\documentclass[onecolumn,draftclsnofoot,12pt]{IEEEtran}
\usepackage{amsfonts}
\usepackage{times}
\usepackage{graphicx}
\usepackage{latexsym}
\usepackage{dsfont}
\usepackage{amssymb}
\usepackage{amsmath}
\usepackage{cite}
\usepackage{verbatim}
\usepackage{subfigure}

\newcommand{\figref}[1]{{Fig.}~\ref{#1}}

\def\bb0{{\mathbb{0}}}

\def\ba{{\mathbf{a}}}
\def\bb{{\mathbf{b}}}

\def\b0{{\mathbf{0}}}

\def\bX{{\mathbf{X}}}

\def\sf0{{\mathsf{0}}}

\usepackage{epstopdf}
\usepackage{enumerate}
\usepackage{algorithmicx}
\usepackage{algorithm}
\usepackage{amsmath}
\usepackage[noend]{algpseudocode}
\usepackage{float}
\usepackage{hyperref}
\usepackage{color}
\usepackage{makeidx}
\usepackage{bbm}
\usepackage{graphicx}
\usepackage{lipsum}
\usepackage{tablefootnote}
\usepackage{multirow}
\usepackage{multicol}
\usepackage{balance}
\usepackage{booktabs}
\usepackage{soul}
\usepackage{xcolor}
\usepackage{autobreak}

\def\j{\mathrm{j}}

\newcommand{\comm}[1]{}

\begin{document}

\title{Vision-Aided 6G Wireless Communications: Blockage Prediction and Proactive Handoff}
\author{Gouranga Charan, Muhammad Alrabeiah, and Ahmed Alkhateeb \footnote{The authors are with the School of Electrical, Computer, and Energy Engineering, Arizona State University. Emails: \{gcharan, malrabei, alkhateeb\}@asu.edu. Part of this work was submitted to IEEE ICC Workshops \cite{charan2020visionaided}.}}

\maketitle

\begin{abstract}

The sensitivity to blockages is a key challenge for the high-frequency (5G millimeter wave and 6G sub-terahertz) wireless networks. Since these networks mainly rely on line-of-sight (LOS) links, sudden link blockages highly threaten the reliability of the networks. Further, when the LOS link is blocked, the network typically needs to hand off the user to another LOS basestation, which may incur critical time latency, especially if a search over a large codebook of narrow beams is needed. A promising way to tackle the reliability and latency challenges lies in enabling proaction in wireless networks. Proaction basically allows the network to anticipate blockages, especially dynamic blockages, and initiate user hand-off beforehand. 
This paper presents a complete machine learning framework for enabling proaction in wireless networks relying on visual data captured, for example, by RGB cameras deployed at the base stations. In particular, the paper proposes a vision-aided wireless communication solution that utilizes bimodal machine learning to perform proactive blockage prediction and user hand-off. 
The bedrock of this solution is a deep learning algorithm that learns from visual and wireless data how to predict incoming blockages. The predictions of this algorithm are used by the wireless network to proactively initiate hand-off decisions and avoid any unnecessary latency. The algorithm is developed on a vision-wireless dataset generated using the ViWi data-generation framework. Experimental results on two basestations with different cameras indicate that the algorithm is capable of accurately detecting incoming blockages more than $\sim 90\%$ of the time. Such blockage prediction ability is directly reflected in the accuracy of proactive hand-off, which also approaches $87\%$. This highlights a promising direction for enabling high reliability and low latency in future wireless networks.

\end{abstract}


\section{Introduction} \label{sec:Intro}
Millimeter-wave (mmWave) and sub-terahertz communications are becoming dominant directions for current and future wireless networks \cite{HeathJr2016, Rappaport2019}. With their large bandwidths, they have the ability to satisfy the high data rate demands of several applications such as wireless Virtual/Augmented Reality (VR/AR) and autonomous driving. Communication in these bands, however, faces several challenges at both the physical and network layers. One of the key challenges stems from the sensitivity of high-frequency signals (i.e., mmWave and sub-terahertz) to blockages \cite{Andrews2016}. These signals suffer from high penetration loss and attenuation, resulting in strong dips in the received Signal-to-Noise Ratio (SNR) whenever an object is present in-between a basestation and a user. Such dips lead to sudden disruptions of the communication channel, which severely impact the reliability of wireless networks. Re-establishing LOS connection is usually done reactively, which brings about a hefty latency burden considering the Ultra-reliable Low-Latency (URLL) requirement of future networks \cite{Bennis2018}. Given all that, high-frequency wireless networks need not only maintain line-of-sight (LOS) connections but also do so proactively, which implies a critical need for a sense of surrounding. 

The aforementioned reliance on LOS draws a striking and important parallel with computer vision, in which visual data (e.g., images and video sequences) only captures visible, i.e., LOS, objects. This parallel is very interesting as computer vision systems rely on machine learning and visible objects to perform a variety of visual tasks depending on object appearance (object detection \cite{Yolo,SSD}) and/or behavior (action recognition \cite{QuoVadis,r-c3d}). In a wireless network, visible objects in the environment are usually the cause of link blockages, and, hence, a computer vision system powered with machine learning could be utilized to provide a much needed sense of surrounding to the network; it enables the network to identify objects in its environment and their behavior and utilize that to \textit{proactively} detect possible blockages. Such capability helps alleviate the strain of link blockages, and as such, this work focuses on developing a \textit{vision-aided dynamic blockage prediction} solution for high-frequency wireless networks. 

\subsection{Prior Works}

The problem of LOS link blockage has long been acknowledged as a critical challenge to high-frequency wireless networks \cite{What5G?,Rappaport2019,Polese2017,Giordani2016}. In those networks, the quality of service highly deteriorates with link blockages. Therefore, solutions centered around multi-connectivity are a major avenue to handle that problem \cite{Polese2017}. For instance, \cite{Giordani2016} proposes a multi-cell measurement reporting system to keep track of the link quality between a mmWave user and multiple basestations. All basestation in that system feed their measurements to a central unit that takes care of cell selection and scheduling. This system is further studied and tested in \cite{Polese2017} under realistic dynamic scenarios. A slightly different look on multi-connectivity is presented in \cite{Aziz2016,Mahmood2019}. In \cite{Aziz2016}, the authors propose a few approaches for multi-connectivity,  all of which focus on utilizing low-frequency bands (sub-6 GHz) to support the mmWave network. \cite{Mahmood2019}, on the other hand, develops a multi-connectivity algorithm that does not only factor in network reliability but also latency. Collectivity, the work on multi-connectivity has its promise and elegance, yet it is lacking on two important fronts. First, it is inherently wasteful in terms of resource utilization; multiple basestations schedule resources for one user as a precaution for probable LOS blockages. The other is its reactive nature; the majority of the multi-connectivity algorithms are designed to react to link blockages, not anticipate them.

A new trend in addressing LOS blockages has been developing in recent years, in which the driving power is machine learning \cite{LSTM_blk,Sub6PredMmWave,Huang2020,BlockagePred}. Some studies such as \cite{LSTM_blk,Sub6PredMmWave} have shown that using wireless sensory data (such as channels and received power), a machine learning model can efficiently differentiate LOS and Non-LOS (NLOS) links. They both address the link blockage problem from a \textit{reactive} perspective, where uni-modal sensory data is first acquired, and then the status of the current link is predicted. The work in \cite{BlockagePred}, however, takes a step forward towards a \textit{proactive} treatment of the problem. It studies proactive blockage prediction and hand-off for a single-moving mmWave user in the presence of \textit{stationary} blockages. The proposed solution utilizes observed sequences of mmWave beamforming vectors (beams) and uses a Gated Recurrent Unit (GRU) network to learn beam patterns that proceed link blockages. Again, despite its appeal, it still falls short in meeting the latency and reliability requirements as the sensory data are only expressive of stationary blockages. On a different note, the work in \cite{CamPredBeam} explores a new dimension for blockage prediction in single user communication settings. It proposes a modified residual network \cite{resnet} that uses visual data to predict stationary blockages. However, like its wireless-data counterparts, it struggles in dealing with complex scenarios with dynamic blockages.

\subsection{Contribution}
In this paper and inspired by the recently proposed Vision-Aided Wireless Communication (VAWC) framework in \cite{CamPredBeam} and \cite{ViWi}, the link-blockage and user hand-off problems are addressed from a proactive perspective. 
Images and video sequences usually speak volumes about the environment they depict, and this is supported by the empirical evidence in \cite{CamPredBeam}. As such, this work develops a deep neural network that learns proactive blockage prediction from sequences of jointly observed mmWave beams and video frames. 
The main contributions of this paper could be summarized in the following few points:

\begin{itemize}
	\item A novel two-component deep learning architecture is proposed to utilize sequences of observed RGB frames and beamforming vectors and learn proactive link-blockage prediction. The architecture harnesses the power of Convolutional Neural Networks (CNNs) \cite{Yolo,SSD} and Gated Recurrent Units (GRU) networks \cite{SpeechRecog, SpeechRecog2}. 
	
	\item The proposed architecture is leveraged to build a proactive hand-off solution. The solution deploys the two-stage architecture in different basestations, where it is used to predict possible future blockages from the perspective of each basestation. Those predictions are streamed to a central unit that determines whether the communication session of a certain user in the environment will need to be handed over to a different basestation or not.   
	
	\item Based on the ViWi data-generation framework \cite{ViWi}, the ViWi-BT challenge scenario \cite{viwi_bt} has been expanded to generate two datasets, namely the blockage-prediction and object-detection datasets. The former is an extension of the ViWi-BT challenge dataset. It provides multi-modal data samples in the form of 4-tuple of image, mmWave beam, link status, and position information, while the latter is a small object detection dataset that provides samples of images, object bounding boxes, and object classes. Both datasets are derived from the ViWi ``ASUDT1\textunderscore28'' scenario \cite{ViWi,ViWi_website}\footnote{Datasets will be made public upon the publication of this work.}.
	
	\item The performance of the two-stage architecture and user hand-off solution are evaluated using a blockage-prediction dataset. The evaluation results confirm the importance of vision-aided blockage prediction in highly dynamic environments; the proposed architecture is shown to be capable of learning proactive blockage prediction from multi-modal data, and it achieves noticeable gain over models that rely solely on wireless data (mmWave beam sequences). 
	
\end{itemize}

The rest of this paper is organized as follows. Section \ref{sec:sys_ch_mod} presents the system and channel models adopted in this work. Section \ref{sec:prob_form} provides a formal description of the link blockage and user hand-off problems addressed in this paper. Section \ref{sec:prop_sol} introduces a detailed description of the proposed solutions for both problems. Sections \ref{sec:exp_set} and \ref{sec:perf_eval}, respectively, present and discuss the description of the experimental setup used to evaluate the performance of the proposed solutions and the main results of that evaluation. Finally, Section \ref{sec:conc} concludes the paper by discussing the main takeaways.

\section{System and Channel Models} \label{sec:sys_ch_mod}
To illustrate the potential of deep learning and VAWC in mitigating the link blockage problem, this work considers a high-frequency communication network where basestations utilize RGB cameras to monitor their environment. The following two subsections provide a detailed description of the system and wireless channel models adopted in this work.

\begin{figure}
	\centering
	\includegraphics[width=0.85\linewidth]{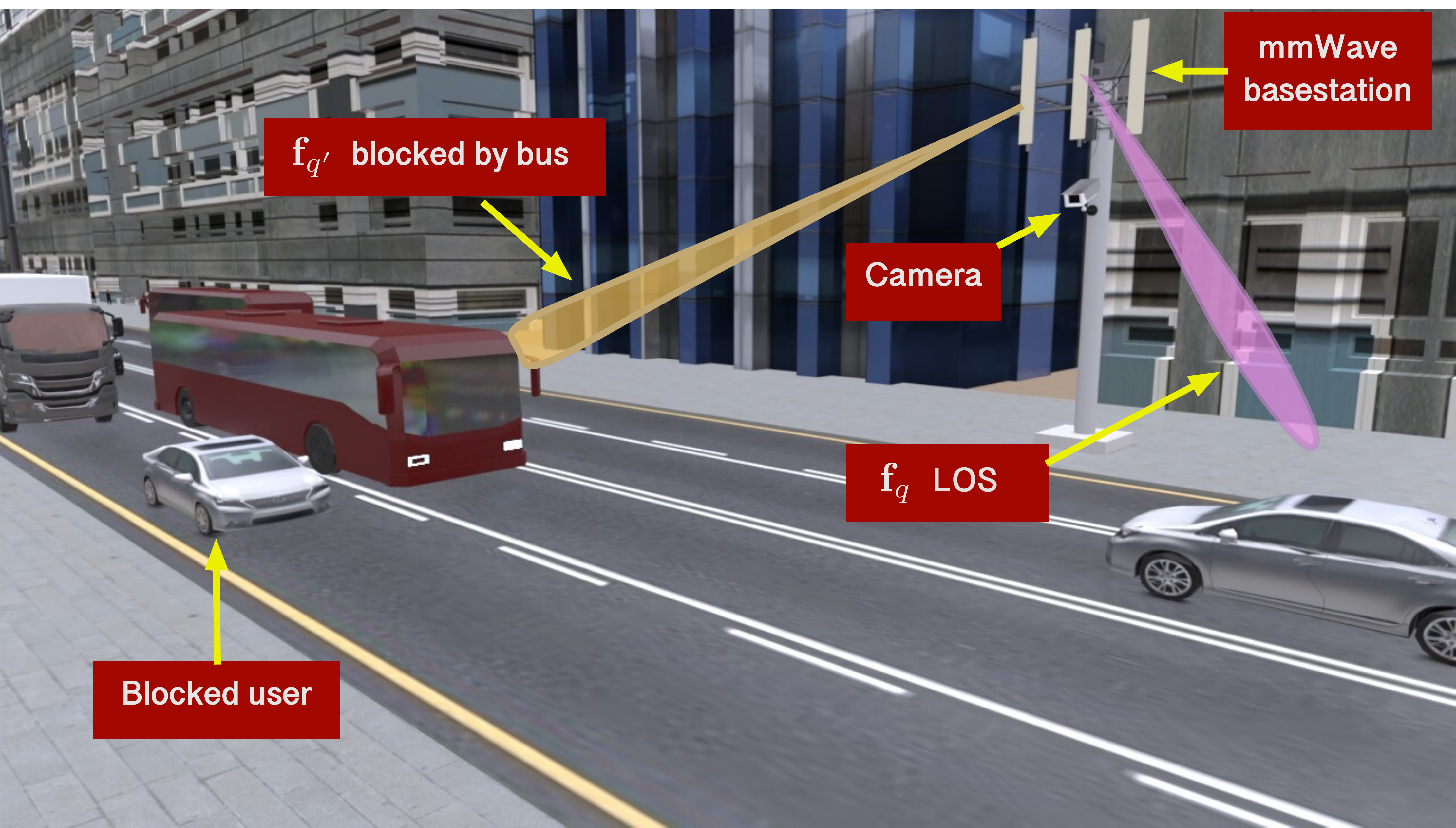}
	\caption{The illustrative figure shows a mmWave basestation equipped with a mmWave array and a camera serving multiple mobile users. One user is LOS while the other is blocked by a bus.}
	\label{fig:sys_mod}
	\vspace{-2mm}
\end{figure}

\subsection{System model}
The communication system considers a small-cell mmWave basestation deployed in an outdoor environment, as depicted in  \figref{fig:sys_mod}. The basestation is equipped with a  uniform linear array (ULA) with $M$ elements and a standard-resolution RGB camera. For practicality \cite{Sub6PredMmWave}, the basestation is assumed to employ analog-only architecture with a single RF chain and $M$ phase shifters. As a result of this architecture, the basestation adopts a predefined beamforming codebook $\boldsymbol{\mathcal F}=\{\mathbf f_q\}_{q=1}^{Q}$, where $\mathbf{f}_q \in \mathbb C^{M\times 1}$ and $Q$ is the total number of beamforming vectors. The choice for $\boldsymbol{\mathcal F}$ in this paper is a beam-steering codebook that follows from the choice of the antenna array, i.e., a ULA. For such codebook, each beamforming vector $\mathbf f_q,\ \forall q\in\{1,\dots,Q\}$ is given by

\begin{equation}\label{eq:array_steering}
	\mathbf f_q = \frac{1}{\sqrt{M}} \left[ 1, e^{j\frac{2\pi}{\lambda}d\sin(\phi_q)},\dots, e^{j(M-1)\frac{2\pi}{\lambda}d\cos(\phi_{q})} \right]^T,
\end{equation}
where $\lambda$ is the wavelength, and $\phi_q \in\{\frac{2\pi q}{Q}\}_{q=0}^{Q-1}$ is a uniform quantization of the azimuth angle with an integer step of $q$. The communication system in this work adopts OFDM with a cyclic prefix of length $D$ and $K$ subcarriers. For any mmWave user in the wireless environment, its received downlink signal is given by
\begin{equation}
    y_{u,k} = \mathbf h_{u,k}^T \mathbf f_q x + n_k,
\end{equation}
where $y_{u,k}\in \mathbb C$ is the received signal of the $u$th user at the $k$th subcarrier, $\mathbf h_{u,k} \in \mathbb C^{M\times 1}$ is the  channel between the BS and the $u$th user at the $k$th subcarrier, $x\in \mathbb C$ is a transmitted complex symbol that satisfies the following constraint $\mathbb E\left[ |x|^2 \right] = P$, where $P$ is a power budge per symbol, and finally $n_k$ is a noise sample drawn from a complex Gaussian distribution $\mathcal N_\mathbb C(0,\sigma^2)$.

\subsection{Channel model}
The channel model adopted throughout this paper is a geometric mmWave channel model with $L$ clusters. This choice of model comes as a result of two facts: (i) the model captures the limited scattering property of the mmWave band \cite{Alkhateeb2015,Alrabeiah_Mapping}, and (ii) the experimental results in this paper are all based on data samples that are partially obtained from a ray tracing simulator as will be describe in Section \ref{sec:exp_set}. The channel vector of the $u$th user at the $k$th subcarrier is given by
 \begin{equation}
\mathbf{h}_{u,k} = \sum_{d=0}^{D-1} \sum_{\ell=1}^L \alpha_\ell e^{- \j \frac{2 \pi k}{K} d} p\left(dT_\mathrm{S} - \tau_\ell\right) \ba\left(\theta_\ell, \phi_\ell\right),
\end{equation} 
where $L$ is number of channel paths, $\alpha_\ell, \tau_\ell, \theta_\ell, \phi_\ell$ are the path gains (including the path-loss), the delay, the azimuth angle of arrival, and the elevation angle of arrival, respectively, of the $\ell$th channel path. $T_\mathrm{S}$ represents the sampling time while $D$ denotes the cyclic prefix length (assuming that the maximum delay is less than $D T_\mathrm{S}$).

\section{Problem formulation} \label{sec:prob_form}
Two significant problems faced in high-frequency wireless networks are LOS link blockages and the ability to perform low-latency user hand-offs. The severity of those two problems mostly revolves around the mixed-dynamics in the wireless environment, i.e., it is characterized by a mixture of dynamic and stationary objects. Developing a solution to the two is tightly linked to equipping the wireless network with a sense of its surroundings; such sense transforms the network from being \textit{reactive} to its environment to being \textit{proactive} in it. This simply means having a network able to predict incoming blockages and initiate hand-off procedures beforehand. With that in mind, this work attempts to utilize machine learning and a fusion of visual and wireless sensory data, e.g., video frames and mmWave beams, to enable that sense of surrounding in a wireless network. The objective is to observe a sequence of a user's image-beam pairs at a basestation and use that sequence to predict whether that user will be blocked within a window of future instances or not. Such prediction task is made possible by two important facts: (i) images or visual data, in general, are rich with information about the scene they depict, e.g., the type of objects, their relative positions to one another, and, in case of videos, the object motion; and (ii) beamforming vectors usually provide directional information that, for well-calibrated antenna arrays, summarizes major signal directions. The following two subsections will lay the groundwork for the proposed solutions by providing formal definitions for the problems of proactive blockage and user hand-off predictions.

\subsection{Blockage Prediction} \label{sec:blockage_pred}
The primary objective of this paper is to utilize sequences of RGB images and beam indices and develop a machine learning model that learns to predict link blockages proactively, i.e., transitions from LOS to NLOS. Formally, this learning problem could be posed as follows. For any user $u$ in the environment, a sequence of image and beam-index\footnote{Since the system model assumes a predefined beamforming codebook, the indices of those beams are used instead of the complex-valued vectors themselves.} pairs is observed over a time interval of $r$ instances. At any time instance $\tau\in \mathbb Z$, that sequence is given by
\begin{equation}
    {\mathcal S}_{u} = \{ (\bX_u[t], b_u[t]) \}_{t = \tau-r+1}^{\tau},
\end{equation}
where $b_{u}[t]$ is the index of the beamforming vector in codebook $\boldsymbol{\mathcal{F}}$ used to serve user $u$ at the $t$th time instance, $\bX_u[t] \in \mathbb{R}^{W \times H \times C}$ is an RGB image of the environment taken at the $t$th time instance, $W$, $H$, and $C$ are respectively the width, height and the number of color channels for the image, and $r\in \mathbb Z$ is the extent of the observation interval. For robust network operation, the objective is to observe ${\mathcal S}_{u}$ and predict whether a blockage will occur within a window of $r^{\prime}\in \mathbb Z$ future instances or not, without focusing on the exact future instance. Let $\mathcal A_{u} = \{a_u[t]\}_{t= \tau+1}^{\tau+r^{\prime}}$ represents the window (sequence) of $r^{\prime}$ future link statuses of the $u$th user, where $a_u[t]\in\{0,1\}$ represents the link status at the $t$th future time instance; and 0 and 1 are, respectively, LOS and NLOS links. Then, the user's future link status $s_u$ in the window $\mathcal A_u$ (henceforth referred to as the \textit{future link status}) could be defined as
\begin{equation}\label{eq:ls}
    s_u = \left\{ \begin{array}{ll}
                  0, & a_u[t] = 0,\ \forall t\in\{\tau+1,\dots,\tau+r^{\prime}\} \\
                  1, & \text{otherwise} \\
                  \end{array}
          \right.
\end{equation}
where $0$ indicates a LOS connection is maintained throughout the window $\mathcal A_u$ and $1$ indicates the occurrence of a link blockage within that window.

The primary objective is attained using a machine learning model. It is developed to learn a prediction function $f_{\Theta}(\mathcal S)$ that takes in the observed image-beam pairs and produces a prediction on the future link status $\hat s \in\{0,1\}$. This function is parameterized by a set $\Theta$ representing the model parameters and learned from a dataset of labeled sequences. To put this in formal terms, let $\mathbb P(\mathcal S, s)$ represent a joint probability distribution governing the relation between the observed sequence of image-beam pairs $\mathcal S$ and the future link status $s$ in some wireless environment, which reflects the probabilistic nature of link blockages in the environment. A dataset of independent pairs $\mathcal D =\{(\mathcal S_u, s_u)\}_{u=1}^{U}$ where $(\mathcal S_u, s_u)$ is sampled at random from $\mathbb P(\mathcal S, s)$---$s_u$ is serving as a \textit{label} for the observed sequence $\mathcal S_u$. This dataset is then used to train the prediction function $f_{\Theta}(\mathcal S)$ such that it maintains high-fidelity predictions for any dataset drawn from $\mathbb P(\mathcal S, s)$. This could be mathematically expressed as
\begin{equation}\label{obj}
\underset{f_{\Theta}(\mathcal S)}{\text{max}} \quad \prod_{u=1}^{U} \mathbb P(\hat s_u = s_u|\mathcal S_u),
\end{equation}
where the joint probability in \eqref{obj} is factored out as a result of the independent and identically distributed samples in $\mathcal D$. This conveys an implicit assumption that for any user $u$ in the environment, the success probability of $f_{\Theta}(\mathcal{S}_u)$ predicting $s_u$ only depends on its observed sequence $\mathcal S_u$.

\subsection{Proactive Hand-off}\label{sec:hand-off}

A direct consequence of proactively predicting blockages is the ability to do proactive user hand-off. In this work the problem is studied for the case of hand-off between two high-frequency basestations, and it is solely based on the availability of a LOS link to a user\footnote{Cases that prompt hand-off like a user approaching the edge of a cell are not considered here.}. Let $\mathcal S_u^{(n)} = \{(\mathbf X^{(n)}_u[t], b^{(n)}_u[t]) \}_{t = \tau-r+1}^{\tau}$ and $\mathcal S_u^{(n^\prime)} = \{(\mathbf X^{(n^\prime)}_u[t], b^{(n^\prime)}_u[t]) \}_{t = \tau-r+1}^{\tau}$ respectively represent the sequences of observed image-beam pairs for the $u$th user at basestation $n$ and $n^\prime$, where $n,n^{\prime}\in\{1,2\}$ such that $n\neq n^{\prime}$. Each of these two sequences is associated with its own future link status, namely $s^{(n)}_u$ and $s^{(n^\prime)}_u$, which are defined similarly to \eqref{eq:ls}. The goal is to determine, with high \textit{confidence}, when a user served by one basestation needs to be handed off to another basestation given the observed two sequence $\mathcal S_u^{(n)}$ and $\mathcal S_u^{(n^\prime)}$. Let $z_u^{nn^{\prime}}\in\{0,1\}$ be a binary random variable indicating whether the $u$th user needs to be handed off from basestation $n$ to basestation $n^{\prime}$ or not, where 1 means a hand-off is needed and 0 means it is not, and let  $\hat{z}_u^{nn^{\prime}} \in \{0,1\}$ be a prediction of the value of $z_u^{nn^{\prime}}$ for user $u$.
  The hand-off confidence is, then, formally described by the conditional probability of successful hand-off, $\mathbb{P}\left(\hat{z}_u^{n n^{\prime}} = z_u^{nn^{\prime}}| \mathcal S_u^{(n)}, \mathcal S_u^{(n^{\prime})} \right)$.

The probability of successful hand-off in the case of two basestations depends on the future link status between a user and those two basestations, and, as such, that probability could be quantified using the predicted and groundtruth future link statuses of the two basestations. Define the tuple of link status predictions $(\hat s_u^{(n)}, \hat s_u^{(n^\prime)}, s_u^{(n)}, s_u^{(n^{\prime})} )$. Then, the event of successful hand-off could be formally expressed as follows  

\begin{equation}\label{eq:HO_events}
	\mathcal H = \mathbbm{1}\{\hat{z}_u^{nn^{\prime}} = z^{nn^{\prime}}\} = \left\{ \begin{array}{l l}
		   1 & , (\hat s_u^{(n)}, \hat s_u^{(n^\prime)}, s_u^{(n)}, s_u^{(n^{\prime})} ) \in \mathcal E_{1}\\
		   0 & , (\hat s_u^{(n)}, \hat s_u^{(n^\prime)}, s_u^{(n)}, s_u^{(n^{\prime})} ) \in \mathcal E_{0}
		  \end{array} \right.
\end{equation}
where $\mathcal E_{0} = \{(0,0,0,0), (0,1,0,1), (1,1,1,1), (1,1,0,0), (1,1,0,1),(0,1,0,0),(0,0,0,1)\}$ indicates the set of tuples (or events) that amount to a successful no hand-off decision while $\mathcal E_{1} = \{(1,0,1,0)\}$ is the set of tuples amounting to a successful hand-off decision. Guided by \eqref{eq:HO_events}, the conditional probability of successful hand-off could be written as 
\begingroup
\allowdisplaybreaks
\begin{align}\label{eq:ho_expan}
&\mathbb{P}\left(\hat{z}_u^{nn^{\prime}} = z^{nn^{\prime}}| \mathcal S_u^{(n)}, \mathcal S_u^{(n^{\prime})} \right) \notag \\
&= \mathbb P( \hat s_u^{(n)}=1, \hat s_u^{(n^\prime)}=0 | s_u^{(n)}=1,s_u^{(n^{\prime})}=0, \mathcal S_u^{(n)}, \mathcal S_u^{(n^\prime)} ) \mathbb P(s_u^{(n)}=1,s_u^{(n^{\prime})}=0) \notag \\
&+ \mathbb P( \hat s_u^{(n)} = 1, \hat s_u^{(n^\prime)} = 1 |s_u^{(n)}=1,s_u^{(n^{\prime})}=1, \mathcal S_u^{(n)}, \mathcal S_u^{(n^\prime)} ) \mathbb P(s_u^{(n)}=1,s_u^{(n^{\prime})}=1) \notag \\
&+ \mathbb P( \hat s_u^{(n)} = 1, \hat s_u^{(n^\prime)} = 1 |s_u^{(n)}=0,s_u^{(n^{\prime})}=0, \mathcal S_u^{(n)}, \mathcal S_u^{(n^\prime)} ) \mathbb P(s_u^{(n)}=1,s_u^{(n^{\prime})}=1) \notag \\
&+ \mathbb P( \hat s_u^{(n)} = 1, \hat s_u^{(n^\prime)} = 1 |s_u^{(n)}=0,s_u^{(n^{\prime})}=1, \mathcal S_u^{(n)}, \mathcal S_u^{(n^\prime)} ) \mathbb P(s_u^{(n)}=0,s_u^{(n^{\prime})}=1) \notag \\
&+ \mathbb P( \hat s_u^{(n)} = 0 | s_u^{(n)}=0, \mathcal S_u^{(n)} ) \mathbb P(s_u^{(n)}=0) 
\end{align}
\endgroup
\eqref{eq:ho_expan} is lower bounded by the probability of joint successful link-status prediction given $\mathcal S_u^{(n)}$ and $\mathcal S_u^{(n^{\prime})}$
\begin{align}\label{eq:ho_low_bound}
\begin{split}
    &\mathbb{P}\left(\hat{z}_u^{nn^{\prime}} = z^{nn^{\prime}}| \mathcal S_u^{(n)}, \mathcal S_u^{(n^{\prime})} \right) \geq \\
	&\sum_{v = 0}^1 \sum_{v^\prime = 0}^1 \mathbb P( \hat s_u^{(n)} = v, \hat s_u^{(n^{\prime})} = v^\prime | s_u^{(n)} = v, s_u^{(n^{\prime})} = v^\prime, \mathcal S_u^{(n)}, \mathcal S_u^{(n^{\prime})} ) \mathbb P(s_u^{(n)}=v, s_u^{(n^{\prime})}=v^{\prime}) \\
    &= \mathbb P(\hat s^{(n)}_u = s^{(n)}_u, \hat s^{(n^{\prime})}_u = s^{(n^{\prime})}_u | \mathcal S^{(n)}_u, \mathcal S^{(n^{\prime})}_u)
\end{split}
\end{align}

Using two blockage-prediction functions $f_{\Theta}(\mathcal S^{(n)})$ and $f_{\Theta}(\mathcal S^{(n^{\prime})})$ (one per basestation), successful proactive hand-off could be viewed from the lens of blockage prediction. More specifically, \eqref{eq:ho_low_bound} indicates that maximizing the conditional probability of joint successful link-status prediction guarantees high-fidelity hand-off prediction. Thus, the two functions $f_{\Theta}(\mathcal S^{(n)})$ and $f_{\Theta}(\mathcal S^{(n^{\prime})})$ need to be learned such that
\begin{equation}\label{eq:pro-handoff}
    \underset{f_{\Theta}(\mathcal S^{(n)}),f_{\Theta}(\mathcal S^{(n^{\prime})})}{\text{max}} \quad \prod_{u=1}^{U} \mathbb P(\hat s^{(n)}_u = s^{(n)}_u, \hat s^{(n^{\prime})}_u = s^{(n^{\prime})}_u | \mathcal S^{(n)}_u, \mathcal S^{(n^{\prime})}_u),
\end{equation}
where $U$ is the total number of samples drawn from the probability distribution $\mathbb P(s_u^{(n)},s_u^{(n^{\prime})},\mathcal S_u^{(n)},\mathcal S_u^{(n^\prime)})$ that governs the relation between the observed sequences $\mathcal S_u^{(n)}$ and $\mathcal S_u^{(n^\prime)}$ and the future link statuses $s_u^{(n)}$ and $s_u^{(n^\prime)}$.

\begin{figure}
    \centering
    \includegraphics[width=1\linewidth]{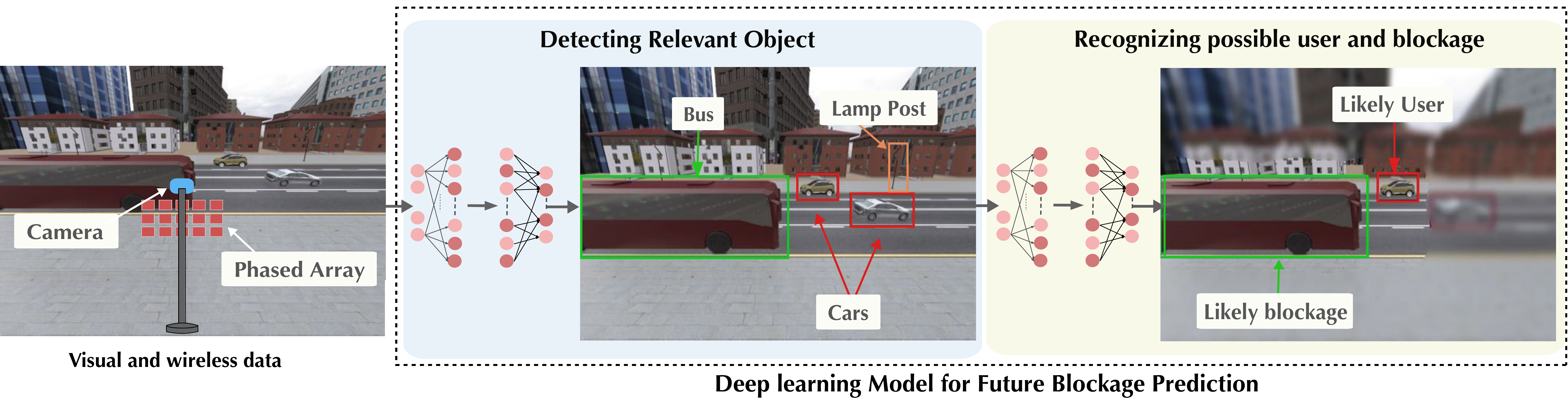}
    \caption{An illustration of the main idea behind the proposed solution. It shows the two notions of detecting relevant objects and zeroing in on those most likely to be the user and its future blockage.}
    \label{fig:key_idea}
    \vspace{-2mm}
\end{figure}

\section{Vision-Aided Dynamic Blockage Prediction and Proactive Handoff} \label{sec:prop_sol}
In this section, we explain our proposed solutions for vision-aided blockage prediction and proactive hand-off using deep learning. The discussion is organized as follows. We first start with highlighting the key idea behind our blockage prediction solution. We further develop that idea by going into the details of our proposed deep learning algorithm. Then, we show how that blockage prediction algorithm is used to address the user hand-off problem.

\subsection{Blockage Prediction: Key Idea}\label{subsec:key_idea}
This work aims to predict future link blockages using deep learning algorithms and a fusion of both vision and wireless data. As we progress from a single-user and stationary blockage \cite{CamPredBeam} to a more realistic scenario with multiple moving objects and dynamic blockages, the task of future blockage prediction becomes far more challenging. A successful prediction of future link blockages in a realistic scene hinges, to a large extent, on the following two notions. First, the ability to detect and identify relevant objects in the wireless environment, objects that could be wireless users or possible link blockages. This includes detecting humans in the scene; different vehicles such as cars, buses, trucks, etc.; and other probable blockages such as trees, lamp posts, \ldots, etc. Second, the ability to zero in on the objects of interest, i.e., the wireless user and its most likely future blockage. Only detecting relevant objects is not sufficient to predict future blockages; it needs to be augmented with the ability to recognize which of those objects is the probable user and which of them is the probable blockage. This recognition narrows the analysis of the scene to the two objects of interest and helps answer the questions of whether and when a blockage will happen. Those two high-level notions are illustrated in \figref{fig:key_idea}.

\begin{figure*}[t]
 	\centering
 	\includegraphics[width=.8\linewidth]{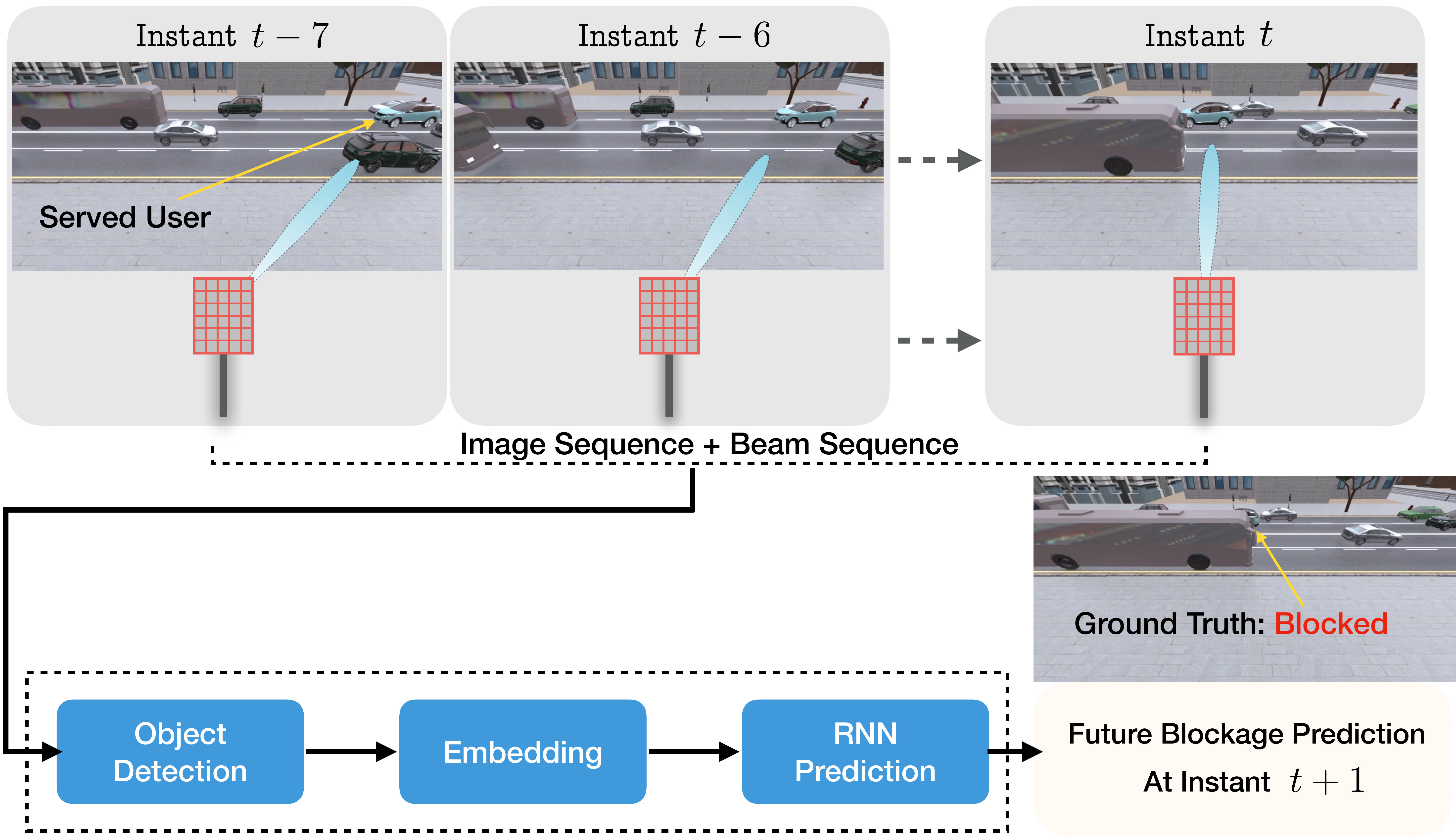}
 	\caption{An overall block diagram illustrating the problem and its solution. The proposed solution takes in a sequence of bimodal image-beam tuples and produces a prediction of the future link status.}
 	\label{fig:concept}
 	\vspace{-2mm}
 \end{figure*}

Guided by the above notions, the prediction function $f_{\Theta}(\mathcal S)$ (or the proposed solution) is designed to break down blockage prediction into two sequential sub-tasks with an intermediate embedding stage. The first sub-task attempts to embody the first notion mentioned above. A machine learning algorithm could detect relevant objects in the scene by relying on visual data alone as it has the needed appearance and motion information to do the job. Given recent advances in deep learning \cite{resnet,vgg,r-c3d}, this sub-task is expected to be well-handled with CNN-based object detectors; they have been setting the bar high for state-of-the-art object detection for about a decade now, see for instance \cite{Yolo,faster_RCNN}. The next sub-task embodies the second notion, recognizing the objects of interest among all the detected objects. Wireless data is brought into the picture for this sub-task. More specifically, mmWave beamforming vectors could be utilized to help with that recognition process. They provide a sense of direction in the 3D space (i.e., the wireless environment), whether it is an actual physical direction for well-calibrated and designed phased arrays or it is a virtual direction for arrays with hardware impairments \cite{CBLNNets}. That sense of direction could be coupled with the set of relevant objects using an embedding stage.
In particular, we propose to observe multiple \textit{bimodal tuples} of beams and relevant objects over a sequence of consecutive time instances, embed each tuple into high-dimensional features, and boil down the second sub-task to a sequence modeling problem. Recurrent neural networks define state-of-the-art \cite{DLBook} for such problems. Hence, we design a recurrent network to implicitly learn the recognition sub-task and produce predictions of future blockages, which is the ultimate goal of our solution. The overall concept of the proposed blockage prediction solution is illustrated in \figref{fig:concept}, and it is going to be detailed in the following couple of sections.

\subsection{Proposed Blockage Prediction Solution}\label{subsec:prop}
\begin{figure*}[t]
	\centering
	\includegraphics[width=\linewidth]{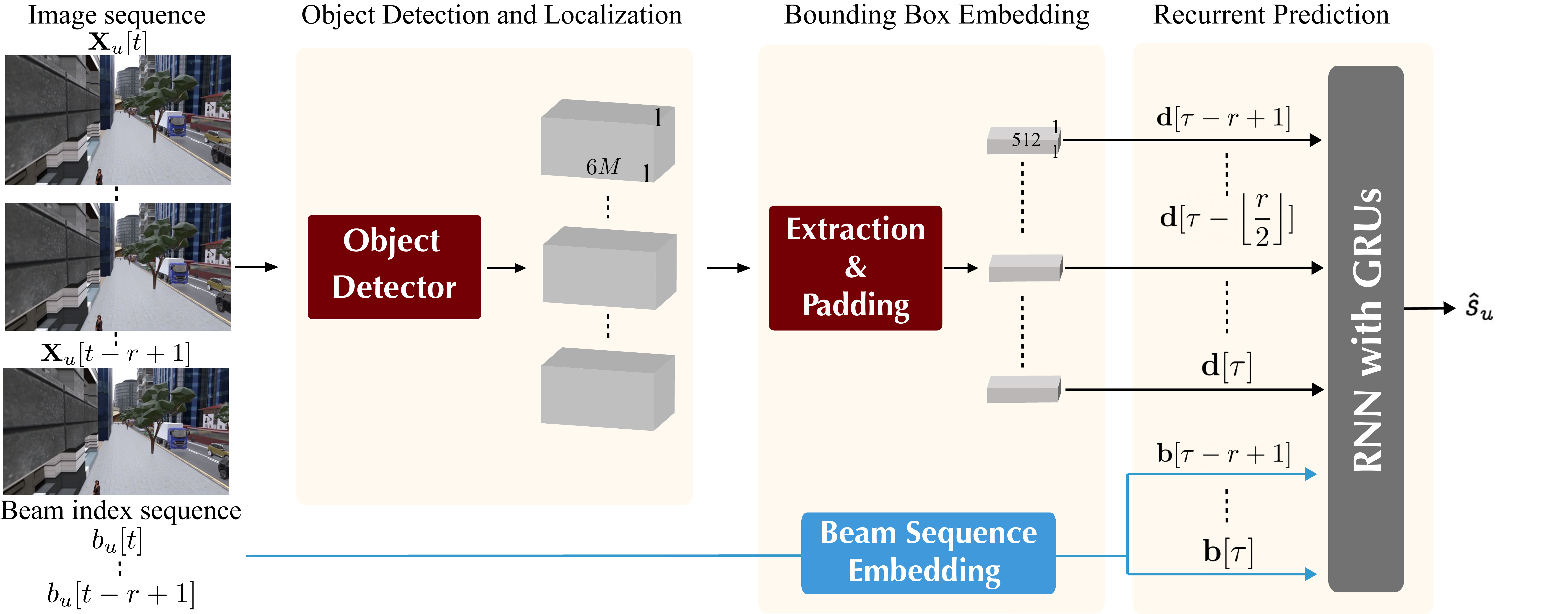}
	\caption{A block diagram showing the proposed neural network. It shows the three main components of the architecture: (i) the object detector, (ii) the embedding component with a CNN network and a beam-embedding block, and (iii) the recurrent prediction network.}
	\label{fig:CNN-RNN}
	\vspace{-2mm}
\end{figure*}

This section takes a deeper dive into the three-component architecture of the machine learning algorithm shown in \figref{fig:concept}. The inner-workings of the architecture are shown in \figref{fig:CNN-RNN} and detailed in the following three subsections.

\subsubsection{Object detector}\label{subsec:obj_det}
The object detector in the proposed solution needs to meet two essential requirements: (i) detecting a wide range of objects and (ii) producing quick and accurate predictions. These two requirements have been addressed in many of the state-of-the-art object detectors. A good example of a fast and accurate object-detection neural network is the You Only Look Once (YOLO) detector, proposed first in\cite{Yolo} and then improved in \cite{Yolov2}. The latest YOLO architecture, YOLOv3, is the best in terms of detection accuracy \cite{Yolov3}, and as such, we adopt it as the object detector in our proposed solution.

\textbf{Choice of object detector:} The YOLOv3 detector is a fast and reliable end-to-end object detection system, targeted for real-time processing. It is a fully convolutional neural network with a feature extraction layer and an output processing layer. Darknet-53 is the backbone feature extractor in YOLO, and the processing output layer is similar to the feature pyramid network (FPN). Darknet-53 comprises 53 convolutional layers, each followed by batch normalization layer and Leaky ReLU activation. In the convolutional layers, 1x1 and 3x3 filters are used. Instead of a conventional pooling layer, convolutional filters with stride 2 are used to downsample the feature maps. This prevents the loss of fine-grained features as the layers learn to downsample the input during training. YOLO makes detection in 3 different scales in order to accommodate different object sizes by using strides of 32, 16, and 8. This method of performing detection at different layers helps address the issue of detecting smaller objects. The features learned by the convolutional layers are passed on to the classifier, which generates the detection prediction. Since the prediction in YOLOv3 is performed using a convolutional layer consisting of 1x1 filters, the output of the network is a feature map consisting of the bounding box co-ordinates, the objectness score, and the class prediction, see \cite{Yolov3}. The list of bounding box co-ordinates consists of top-left co-ordinates and the height and width of the bounding box. In this work, we compute the center co-ordinates of each of the bounding boxes from the top-left and the height and the width of the box. 

\textbf{Integration of object detector:} Instead of building and training the YOLOv3 from scratch, the proposed solution utilizes a pre-trained YOLO network and integrates it into its architecture with some minor modifications. First, the network architecture is modified to detect the objects of interest, e.g., cars, buses, trucks, trees, \ldots, etc; the number of those objects and their types (classes) are selected based on the target wireless environment in which the proposed solution is going to be deployed. For any choice of the number of objects and classes, the modification on the YOLOv3 architecture only affects the size of the classifier layer, which allows us to take advantage of the other trained layers. Second, the YOLOv3 network with the modified classifier is then fine-tuned using a dataset resembling the target wireless environment. This step adjusts the classifier and, as the name suggests, fine-tune the rest of the architecture to be more attentive to the objects of interest.

\subsubsection{Bounding Box Extraction and Beam Embedding}
\label{subsec:feat_beam_emb}
The prediction function relies on dual-modality observed data, i.e., visual and wireless data. Although such data is expected to be rife with information, its dual nature brings about a heightened level of difficulty from the learning perspective. In an effort to overcome that difficulty, the proposed solution incorporates an embedding component that processes the extracted bounding box values and the beam indices separately, as shown in the embedding component of Fig.\ref{fig:CNN-RNN}. It transforms them to the same $N$-dimensional space before they are fed to the next component. For beam indices in the input sequence, the adopted embedding is simple and does not require any training. It generates a lookup table of $\left|\boldsymbol{\mathcal F}\right|$ real-valued vectors $\mathbf b[t]\in \mathbb R^N$ where $t\in\{\tau-r+1, \dots, \tau\}$. The elements of each vector are randomly drawn from a Gaussian distribution with zeros mean and unity standard deviation.

For bounding boxes output by the object detector, they undergo a simple transform-and-stack operation. In particular, each bounding box is transformed into a 6-dimensional vector comprising the center co-ordinates $[x_{\text{cent}}, y_{\text{cent}}]$, the bottom left co-ordinates $[x_1, y_1]$, and the top right co-ordinates $[x_2, y_2]$. The co-ordinates are normalized to fall in the interval $[0,1]$. They, collectively, help in marking the exact location of an object in the scene. Then, the transformed bounding boxes of one image (or video frame) are stacked to form one high-deminsional vector $\tilde{\mathbf d}[t]\in\mathbb R^{6M \times 1}$, where $M$ is the number of objects detected in an image and $t\in\{\tau-r+1, \dots, \tau\}$. Since the solution is proposed for dynamic wireless environments, the number of objects in each image is not fixed, resulting in a variable-length $\tilde{\mathbf d}[t]$. Therefore, $\tilde{\mathbf d}[t]$ is padded by $N-M$ zeros to transform it into a fixed length vector $\mathbf d[t]\in\mathbb R^{N\times 1}$.

\subsubsection{Recurrent prediction}
\label{subsec:rnn}
CNN networks inherently fail in capturing sequential dependencies in input data; thereby, they are not expected to learn the relation within a sequence of embedded features. To overcome that, the third component of the proposed architecture utilizes Recurrent Neural Networks (RNNs) and performs future blockage prediction based on the learned relation among those features. In particular, the recurrent component has two layers of Gated Recurrent Units (GRU) separated by a dropout layer. These two layers are followed by a fully-connected layer that acts as a classifier. The recurrent component receives a sequence of length $2r$ of bounding-box and beam embeddings, i.e., a sequence of the form $\{\mathbf d[\tau-r+1], \dots, \mathbf d[\tau],  \mathbf b[\tau-r+1],\dots, \mathbf b[\tau]\}$. Hence, it implements $2r$ GRUs per layer. The output of the last unit in the second GRU layer is fed to the classifier to predict the future link status $\hat s_u$.

\subsection{Proactive Hand-off}\label{sec:hand-off_sol}
A major advantage of proactive blockage prediction is that it enables mitigation measures for LOS-link blockages in small-cell high-frequency wireless networks, such as proactive user hand-off. The predictions of a vision-aided blockage prediction algorithm could serve as a means to anticipate blockages and re-assign users based on LOS link availability. To illustrate that, the deep learning architecture presented in Section \ref{subsec:prop} is deployed in a simple network setting, which embodies the setting adopted in Section \ref{sec:prob_form}\footnote{It is important to note that: (i) extension of this proposed solution to more than two basestation is straightforward, and (ii) the two basestation example is used for clarity.}. Two adjacent small-cell high-frequency basestations are assumed to operate in the same wireless environment. They are both equipped with RGB cameras that monitor the environment, and they are also running two copies of the proposed deep architecture. A common central unit is assumed to control both basestations and have access to their running deep architectures. Each user in the environment is connected to both basestations but is only served by one of them at any time, i.e., both basestations keep a record of the user's best beamforming vector at any coherence time, but only one of them is servicing that user. The objective in this setting is to learn two blockage-prediction functions and use their predictions in the central unit to perform proactive user hand-off. More formally, we aim to learn the two prediction functions $f_{\Theta}(\mathcal S^{(n)})$ and $f_{\Theta}(\mathcal S^{(n^{\prime})})$ that could maximize \eqref{eq:pro-handoff}.

\textbf{Proposed user hand-off solution:} From \eqref{eq:pro-handoff}, functions $f_{\Theta}(\mathcal S^{(n)})$ and $f_{\Theta}(\mathcal S^{(n^{\prime})})$ need to maximize the joint conditional probability of successful link-blockage prediction. Such requirement, albeit being accurate, may not be computationally practical as it requires a joint learning process, which may not scale well in an environment with multiple small-cell basestations. Thus, we propose to train two independent copies of the blockage prediction architecture on two separate datasets, each of which is collected by one basestation. This choice could be formally translated into a conditional independence assumption in \eqref{eq:pro-handoff}. More specifically, for the $u$th user, the event of successful link-status prediction at basestation $n$, i.e., $\{\hat s^{(n)}_u = s_u^{(n)}|\mathcal S_u^{(n)}\}$, is independent from that of the same user at basestation $n^{\prime}$, i.e., $\{\hat s^{(n^\prime)}_u = s_u^{(n^\prime)}|\mathcal S_u^{(n^\prime)}\}$. The intuition behind this assumption is rooted in the camera orientation at each basestation; each camera could view the environment from a different view-angle, which could result in different object positions, object orientations, motion directions, and image background. The trained deep architectures are deployed once they reach some satisfying generalization performance. At any time instance, the two architectures feed their predictions to the central unit, and the unit uses them to anticipate whether a user should be handed off or not (i.e., $z^{nn^{\prime}} = 1\ \forall n,n^{\prime}\in\{1,2\}$ and $n \neq n^{\prime}$). A hand-off is only initiated when the LOS link at the serving basestation is predicted to be blocked while the LOS link at the other basestation is predicted to be maintained.

\section{Experimental Setup}\label{sec:exp_set}
In order to evaluate the proposed solution for both blockage prediction and proactive hand-off, this section introduces a discussion on the communication scenario considered for the evaluation experiments, the process of generating the development and evaluation datasets, the evaluation metrics used to assess the performance, and the training procedure of the proposed two-stage neural network. 

\begin{figure}[t]
    \centering
    \subfigure{
    	\centering
    	\includegraphics[width=.489\textwidth]{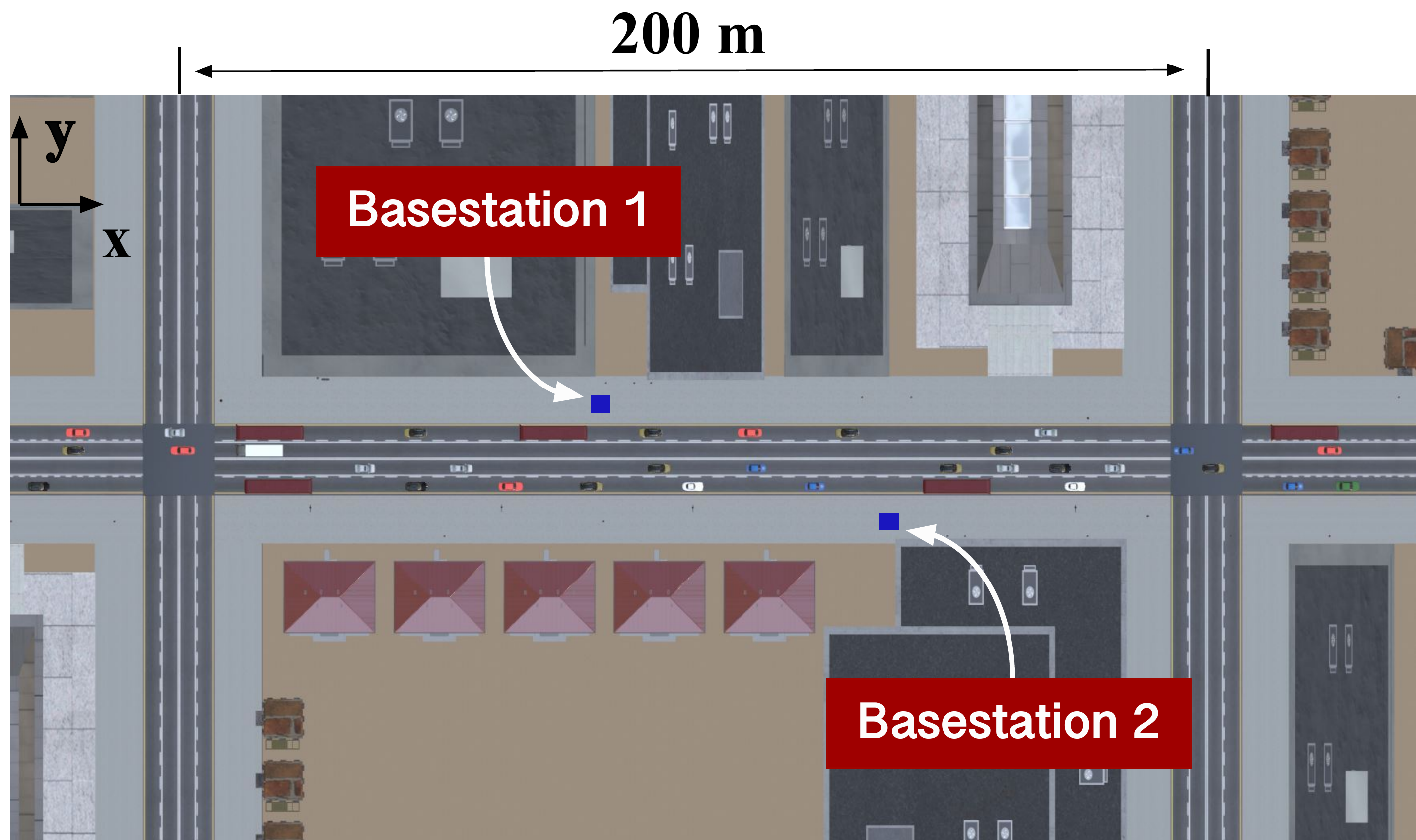}}
	\subfigure{
    	\centering
    	\includegraphics[width=.475\textwidth]{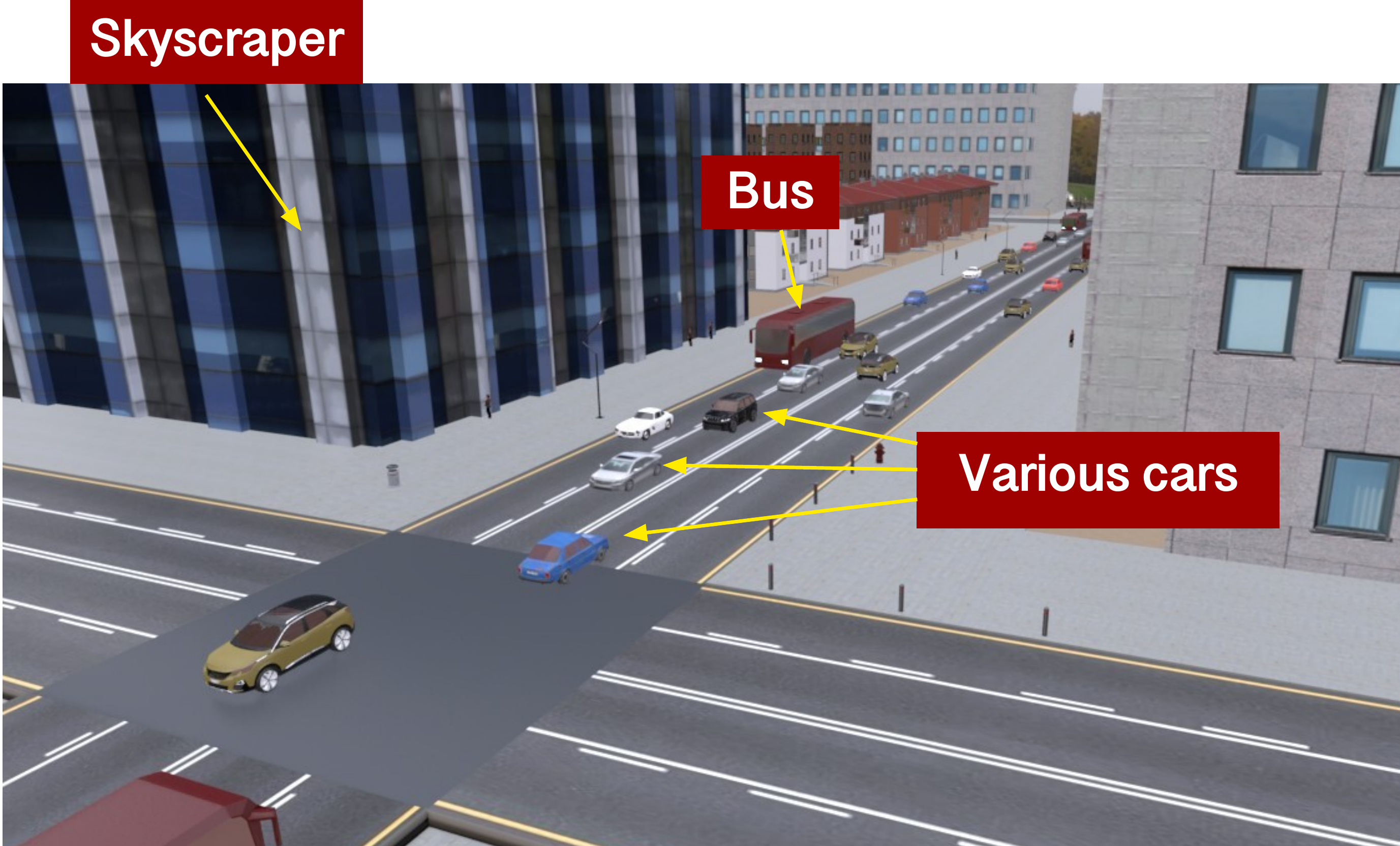}
}

    \caption{A top-view (a) and a perspective view (b) of the simulation outdoor scenario. It is modeled after a busy downtown street with a variety of moving and stationary objects, such as cars, buses, trucks, pedestrians, high-rises, etc. The view also depicts the two basestations.}
    \label{fig:top_view}
    \vspace{-4mm}
\end{figure}

\subsection{Communication Scenario and Datasets}
We first start by describing the communication scenario used for our development and evaluation experiments. Then, we give a detailed description of the two datasets generated from that scenario.  

\subsubsection{Scenario description}
The scenario considered in this paper is the ViWi multi-user scenario ``ASUDT1\textunderscore28'' \cite{ViWi,ViWi_website}, which is an outdoor mmWave communication environment built using a game engine and a ray-tracing software. It is developed using the ViWi data generation framework. The scenario depicts a typical downtown street with its various elements, vehicles, pedestrians, lamp posts, skyscrapers, \ldots, etc, see \figref{fig:top_view}. Each vehicle on the street represents a possible user, and large vehicles, like buses and trucks, act as dynamic blockages to smaller ones, i.e., cars. The scenario has a total of 60 vehicles, 2 trucks, 8 buses, and 50 cars, and they all move at different speeds. The scenario also considers two small-cell basestations operating at 28 GHz. They are set 80 m away from each other and on opposite sides of the street, as \figref{fig:top_view} shows. Each one is 4.5 m above the ground and equipped with three differently-oriented cameras providing two side views and a central view of the street. Using the ViWi data-generation script, a raw dataset of 4-tuples is generated where each of those tuples has co-existing vision-wireless information concerning one user at a certain time instance in the environment, i.e., for a user $u$ and time instance $t$, a 4-tuple consists of an image, mmWave channels, link status, and location. This raw dataset is henceforth referred to as the \textit{seed dataset}. It is further processed to obtained two development and evaluation datasets. The first is for training and testing the object detector, while the other is used to train and evaluate the proposed deep neural network used to address blockage prediction and user hand-off problems. The following two subsections will shed more light on each one of these two datasets.

\subsubsection{Tiny Object Detection Dataset}
\label{sec:manual_dataset}
A dataset of $600$ samples is selected from the seed dataset generated using ViWi ASUDT1\textunderscore28 scenario to fine-tune the COCO pre-trained YOLOv3. As mentioned above, there are a total of six cameras, three for each basestation. Each of these cameras covers a different portion of the street and views the objects from a different orientation. A total of $300$ image samples are selected from the central cameras, and the remaining $300$ samples come from the side cameras. This is done to incorporate the difference in orientations, resulting in a diverse dataset. The ViWi ASUDT1\textunderscore28 scenario is not object detection ready, i.e., it does not contain the labels and the bounding boxes of the objects present in the scene. Therefore, the $600$ samples are manually labeled to create the object-detector dataset. This dataset has bounding-box labels of various cars, trucks, and buses. It is split $80\%-20\%$ to create training and validation sets.

\subsubsection{Blockage Prediction Dataset}\label{sec:blk_dataset}

To generate a blockage and user hand-off dataset, the seed dataset undergoes a processing pipeline that eventually generates a dataset with observed sequences and label pairs, i.e., $\mathcal D$. Recognizing that the proposed architecture requires sequences of image-beam inputs, the first step in the pipeline is to generate beamforming vectors from the mmWave channels in the seed dataset. The result is a dataset equal in size to the seed dataset but with beamforming vectors instead of mmWave channels. Then, the second step in the pipeline creates the input sequences and their corresponding labels. For every user in the environment, every 13 consecutive 3-tuples of image, beam, and link status are stacked to form one raw sequence. In that sequence, the first 8 images (i.e., $r = 8$) and beams are paired to form the observed sequence $\mathcal S_u^{(\beta)}$ where $\beta\in\{n, n^{\prime}\}$. The last 5 (i.e., $r^{\prime} = 5$) link statuses in the raw sequence are used to construct the label of the observed sequence as described by \eqref{eq:ls}. The final result of the second pipeline step is a large collection of observed sequences and their labels, a little shy of 2 million sequences. 

That large collection is composed of multiple cases in terms of labels, the majority of which have a LOS future link status (i.e., $s_u = 0$). Thus, the third pipeline step attempts to reduce the size of the dataset and balance out its labels. This is done by randomly and equally sampling observed sequences. More specifically, approximately $27000$ observed sequences with LOS labels (henceforth referred to non-pivotal sequences) and another approximately $27000$ with NLOS labels (henceforth referred to the pivotal sequences) are randomly sampled from the large collection. These sampled sequences are divided equally among the 6 cameras (2 basestations) in the scenario, resulting in $9000$ pivotal and non-pivotal sequences per camera. The final outcome of the third step is the blockage-prediction and user hand-off dataset, i.e., $\mathcal D = \{(\mathcal S_u, s_u)\}_{u = 1}^{U}$ where $U = 54000$. This dataset is split $50\%-50\%$ into training and validation sets. 

\subsection{Evaluation Metrics}
The success of future blockage prediction is heavily dependent on the performance of the object detector. Correctly detecting the objects in the scene, especially the user, and the blockage is of paramount importance. The evaluation metric for the object detection task is different from that of blockage prediction. Therefore, in this subsection, we present the metric used for evaluating the success of both the object detection task and the future link blockage prediction task separately.

\subsubsection{Object Detection Evaluation Metric}\label{sec:mAP}
The objectives of an object detection model are two folds: (i) \textbf{Classification:} The object detection model needs to identify the objects present in the image and respective classes of the object. (ii) \textbf{Localization:} The object detection model also needs to predict the bounding boxes around each detected object, thereby detecting the location of the objects in the image. Therefore, it is necessary to evaluate the performance of both the classification of different objects and the localization of using bounding boxes in the image.  

Several mathematical tools such as the Intersection over Union (IoU), and confidence score are used to quantify the quality of the object detector outputs. The confidence score is the probability that a bounding-box predicted by the model contains an object. The IoU computes the area of the intersection divided by the area of the union of the two bounding boxes, i.e., the ground truth ($\mathcal{B}_g$) and the predicted bounding box ($\mathcal{B}_g$). IoU can be defined as
\begin{equation}
    \text{IoU} = \frac{\text{area}(\mathcal{B}_g \cap \mathcal{B}_p)}{\text{area}(\mathcal{B}_g \cup \mathcal{B}_p)}.
\end{equation}
Both the IoU and the confidence score are utilized to evaluate the detection performance, more specifically calculate precision, recall, average precision (AP), and mean average precision (mAP) scores of the model. For more details on those metrics, the reader is referred to \cite{ImageNet12,PASCAL}.

\subsubsection{Blockage Prediction Evaluation Metric}
In this section, we present the evaluation metric followed to evaluate the performance of our proposed blockage prediction solution. Post-training, we evaluate each network on the validation set. The primary method of evaluating the model performance is using top-1 accuracy, which is defined as follows
\begin{equation}
    Acc_{top-1} = \frac{1}{U^\prime} \sum_{u=1}^{U^\prime} \mathbbm{1} \{\hat s_u = s_u \},
\end{equation}
where $\hat s_u$ is the predicted blockage value for user $u$ when provided with the sequence of observation $S_{u}$ as defined in Section~\ref{sec:prob_form}, $s_u$ is the ground-truth value of the same data sample, $U^\prime$ is the total number of data samples in the validation dataset, $\mathbbm{1}\{.\}$ is the indicator function, with the value of 1 only when the condition provided is satisfied. We also resort to precision and recall for a more detailed look into the blockage-prediction performance.


\begin{table}[!t]
	\caption{Design and Training Hyper-parameters}
	\label{table}
	\centering
	\setlength{\tabcolsep}{5pt}
	\renewcommand{\arraystretch}{1.2}
	\begin{tabular}{|l|c|c|}
		\hline
		\multirow{5}{*}{Design}   
		
		& Number of GRUs Per Layer ($2r$)   & $16$             \\  \cline{2-3}
		& Embedding Dimension ($N$)        & $256$                        \\ \cline{2-3} 
		& Hidden State Dimension         & $64$                         \\ \cline{2-3} 
		& Number of classes  & $2$                        \\  \cline{2-3}  
		& Dropout Percentage             & 0.3                        \\ \hline \hline
		\multirow{4}{*}{Training} & Optimizer                      & ADAM                       \\ \cline{2-3} 
		& Learning Rate                  & $1 \times 10 ^{-3}$ \\ \cline{2-3} 
		& Batch Size                     & $200$                       \\ \cline{2-3} 
		& Number of Training Epochs               & $100$                         \\ \hline
	\end{tabular}
	\label{tab_params}
	\vspace{-4mm}
\end{table}


\subsection{Network Training}
In this subsection, we present the training methodology of both the object detector network and the proposed blockage prediction solution. All the experiments were performed on a single NVIDIA RTX Titan GPU using the PyTorch deep learning framework. 

\subsubsection{Fine-tuning object detector}\label{sec:finetune_yolo}
The parameters of YOLOv3 are pre-trained on the COCO dataset \cite{coco}, which is a large-scale image dataset with around 80 object classes. However, there are mainly $3$ objects of interest in the blockage-prediction dataset as mentioned in Section~\ref{sec:manual_dataset}. The COCO pre-trained YOLOv3 object detector model is fine-tuned on the tiny object-detection dataset to improve its performance in detecting the objects of interest. In order to achieve a robust object detection model, separate models are trained for each camera in the scenario. The YOLOv3 detector is trained with Adam optimizer. The bounding boxes are extracted with a confidence threshold of $0.5$ and an NMS threshold of $0.45$.

\subsubsection{Training the blockage-prediction architecture}

\begin{figure}[t]
    \centering
    \subfigure[Input Image to YOLOv3] {
    	\centering
    	\includegraphics[width=.6\textwidth]{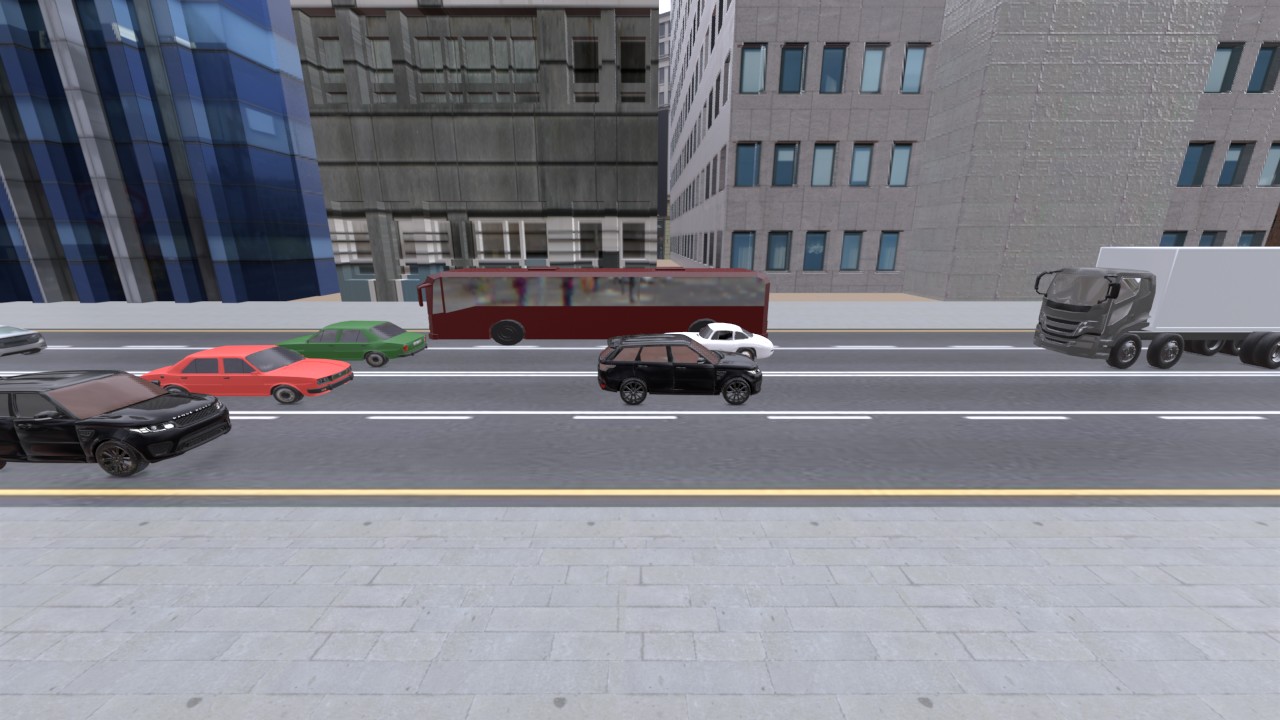}
}
    \subfigure[Output from the pre-trained detector] {
    	\centering
    	\includegraphics[width=.47\textwidth]{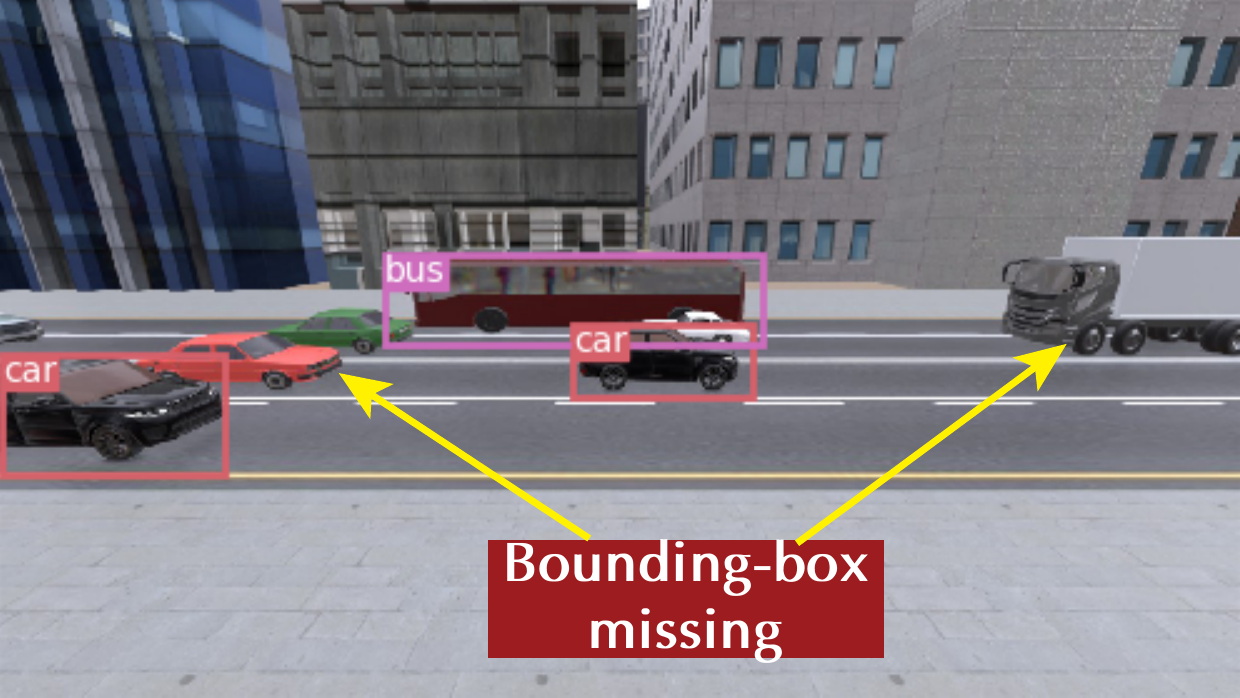}
}
    \subfigure[Output of the fine-tuned detector]{
    	\centering
    	\includegraphics[width=.47\textwidth]{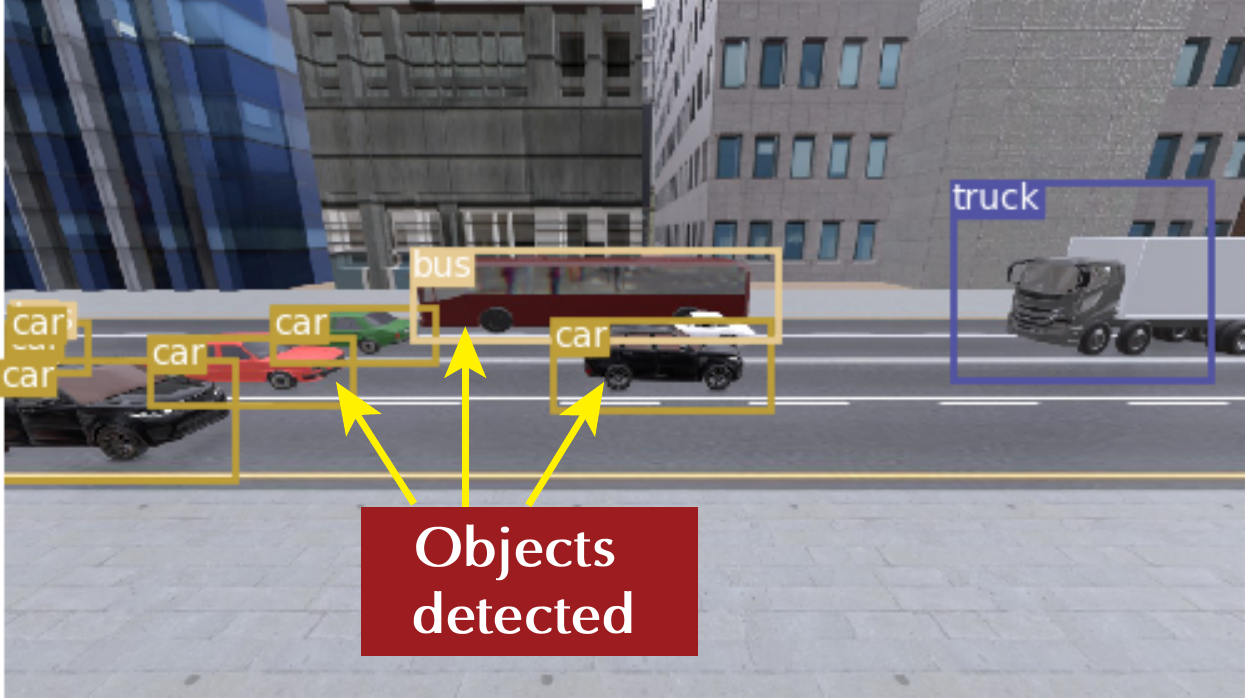}
}
    \caption{A visualization of the output from YOLOv3 object detector. (a) Output from the pre-trained detector; (b) Output of the fine-tuned detector. In this example, the user and the blockage are the red bus and the truck in the fourth and third lanes. It is observed that the fine-tuned model clearly improves the object detection quality compared to the pre-trained one.}
    \label{ref:yolo_detect}
    \vspace{-4mm}
\end{figure}

This paper studies the ability of the proposed deep architecture to perform blockage prediction using sequences of RGB images and observed beams. The blockage prediction dataset described in Section.\ref{sec:blk_dataset} is used to train the second component of the architecture, recurrent prediction, as the first component is kept fixed following the fine-tuning of the object detector. The training hyper-parameters are listed in Table~\ref{tab_params}, and a cross-entropy loss is used to guide the training process \cite{DLBook}. To highlight the potential of vision-aided link-blockage prediction, we develop a baseline model that performs the same task but without the visual data, using only the beam sequences. The model is simply the recurrent component of the proposed solution described in Section~\ref{subsec:rnn}. It takes in the 8-beam sequences and predicts the link-status. The training of this baseline solution is also performed with a cross-entropy loss. 

\section{Performance Evaluation} \label{sec:perf_eval}
In this section, we first analyze the performance of the fine-tuned YOLOv3 object detector before delving into analyzing the performance of our proposed solution for future blockage prediction and proactive hand-off.


\begin{table}[!t]
	\caption{Comparison of the AP precision score of COCO pre-trained and ViWi fine-tuned YOLOv3 model}
	\label{table}
	\centering
	\setlength{\tabcolsep}{5pt}
	\renewcommand{\arraystretch}{1.2}
	\begin{tabular}{|c|c|c|c|c|c|c|}
		\hline
		\multirow{2}{*}{\textbf{YOLOv3}}                & \multirow{2}{*}{\textbf{\begin{tabular}[c]{@{}c@{}}Confidence\\ Threshold\end{tabular}}} & \multirow{2}{*}{\textbf{\begin{tabular}[c]{@{}c@{}}NMS \\ Threshold\end{tabular}}} & \multicolumn{3}{c|}{\textbf{AP}}                    & \multirow{2}{*}{\textbf{mAP}} \\ \cline{4-6}
		&                                                                                          &                                                                                    & \textbf{Car}    & \textbf{Truck}  & \textbf{Bus}    &                               \\ \hline \hline
		\textbf{COCO Pre-Trained Model}                 & 0.5                                                                                      & \multirow{4}{*}{0.6}                                                               & 0.0079          & 0.0             & 0.6486          & 0.1566                        \\ \cline{1-2} \cline{4-7} 
		\multirow{3}{*}{\textbf{ViWi Fine-Tuned Model}} & \textbf{0.6}                                                                             &                                                                                    & \textbf{0.8444} & \textbf{0.8968} & \textbf{0.9929} & \textbf{0.9114}               \\ \cline{2-2} \cline{4-7} 
		& 0.8                                                                                      &                                                                                    & 0.8290          & 0.8968          & 0.9929          & 0.9062                        \\ \cline{2-2} \cline{4-7} 
		& 0.9                                                                                      &                                                                                    & 0.8034          & 0.8968          & 0.9353          & 0.8785                        \\ \hline
	\end{tabular}
	\label{tab_mAP}
	\vspace{-4mm}
\end{table}


\subsection{Object Detector Performance}

The object detector plays a crucial role in the proposed architecture; it performs the first sub-task of identifying the relevant objects in a wireless environment. The correct prediction of future blockage is heavily dependent on the performance of this object detector. Although the YOLOv3 pre-trained on the COCO dataset can detect the objects of interest in our dataset, i.e., the cars, buses, and trucks, its performance is relatively poor on the blockage-prediction dataset. Therefore, as proposed in Section~\ref{sec:finetune_yolo}, the pre-trained YOLOv3 is further fine-tuned on the tiny object-detector dataset. To show the difference in performance, Fig.~\ref{ref:yolo_detect} depicts an example output of both the COCO pre-trained YOLOv3 and the fine-tuned YOLOv3 model. \textbf{In this specific scene, the user is the red bus, and the blockage is the truck in the third lane.} This particular example is very interesting for our analysis as the user is about the get blocked by the truck in the future. The pre-trained YOLOv3 model fails to detect both the blockage as observed in Fig.~\ref{ref:yolo_detect}(b), which the fine-tuned model overcomes. This failure will be propagated downstream to the next component in the proposed architecture, and it is likely to lead to a wrong blockage prediction as the recurrent component is completely oblivious to the presence of the user and its blockage.

\begin{figure}[t]
	\centering
	\subfigure[Beam-only Approach]{	\includegraphics[width=.4\columnwidth]{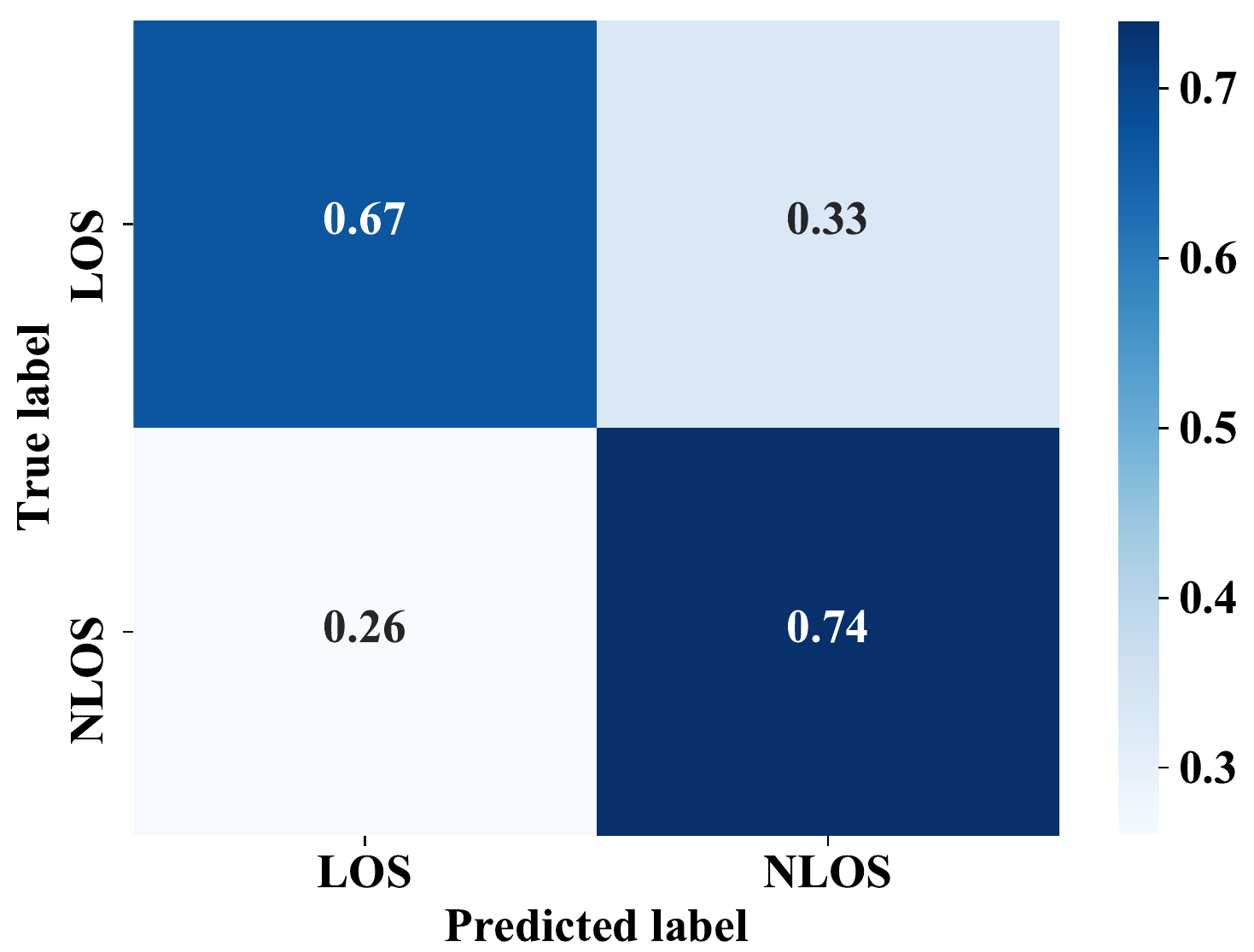}	\label{fig:beam_only}}
	\subfigure[Proposed Vision-Aided Approach]{	\includegraphics[width=.4\columnwidth]{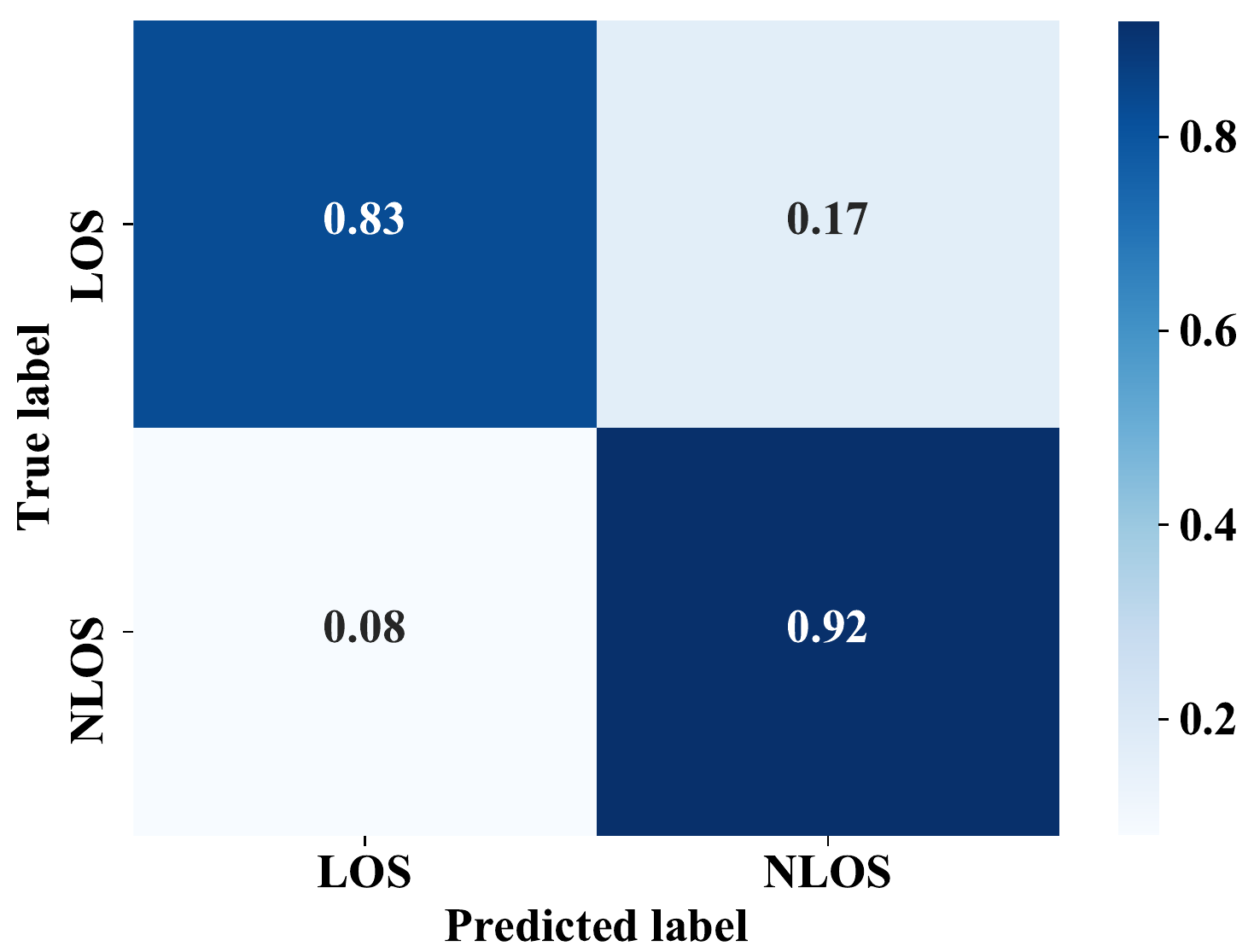}\label{fig:beam_image}}
	\caption{The confusion matrices for the task of future blockage prediction, for both the (a) baseline (beam-only) solution and (b) the proposed vision-aided deep learning approach. The proposed solution achieves a precision and recall of 84\% and 92\% respectively, as compared to 67\% and 74\% of the baseline (beam-only) solution. This highlights the accurate prediction capability of the proposed vision-aided approach. }
	\label{fig:conf_mats}
	\vspace{-4mm}
\end{figure}

To quantify the performance of the object detector, we calculate the AP and the mAP on the validation dataset. The YOLOv3 model is separately trained and validated for the images coming from the central and side cameras. This is done to ensure a robust object detection and classification performance. Table~\ref{tab_mAP} shows the average AP of the pre-trained and fine-tuned models on all $3$ classes of relevant objects, i.e., car, bus, and truck, across all the cameras. The performance of the fine-tuned model surpasses that of the pre-trained one. The fine-tuned YOLOv3 model achieves approximately $\textbf{6} \times$ the mAP performance of the pre-trained model. It is likely that the pre-trained model performance drops on the validation dataset because there is a shift in the data distribution between COCO and the ViWi dataset. Such improvement in the detector performance significantly reflects on the overall blockage prediction and proactive hand-off capabilities of our proposed solution. This is discussed in the next few sections.

\subsection{Blockage Prediction}
In this subsection, we present a detailed analysis of our proposed deep architecture for future blockage prediction. As mentioned in Section~\ref{sec:blk_dataset}, there are $2$ basestations, each equipped with $3$ cameras pointing in different directions, covering the whole street. Each of the three cameras presents a different set of challenges to the prediction task. In order to develop a thorough understanding of the proposed architecture, it is trained and tested on the data samples of each camera separately. The results of all those tests are presented and discussed in the following hierarchy. We first analyze the overall performance on all validation samples (i.e., all cameras). Then, we narrow down the scope of the discussion to the camera-wise performance, and we conclude this section with a discussion on the performance of the proposed architecture on the pivotal sequences.

\subsubsection{Overall Analysis}
As mentioned above, $6$ copies of the proposed architecture are trained, one for each camera.
 In \figref{fig:conf_mats}, we present the confusion matrices for both the baseline and the proposed solutions. Given an almost balanced validation set, i.e., a total of $4500$ samples per camera for each label, the figure shows that beamforming vectors alone do not reveal enough information about future blockages; the baseline solution achieves around $70.48\%$ top-1 accuracy, leaving significant room for improvement. The proposed solution, on the other hand, achieves approximately $17\%$ improvement in accuracy over that of the baseline solution. This highlights the importance of visual data in tackling the link-blockage prediction problem. Accuracy is a holistic measure that may not reflect the intricacies of blockage prediction. Hence, we take a closer look at the performance by evaluating the precision-recall performance. For the proposed deep architecture, it achieves a precision of $84\%$ at a recall rate of $92\%$. This reflects a high level of trustworthiness in the blockage predictions of the architecture; it successfully identifies $92\%$ of all blockage cases while maintaining $84\%$ accurate predictions in mixed LOS and NLOS cases. In contrast, the precision of the baseline model falls to $69\%$ at a recall of only $74\%$, which re-affirms the value of vision for proactive blockage prediction.

\begin{figure}[t]
	\centering
	\subfigure[Camera 3 of Basestation 1]{	\includegraphics[width=0.48\columnwidth]{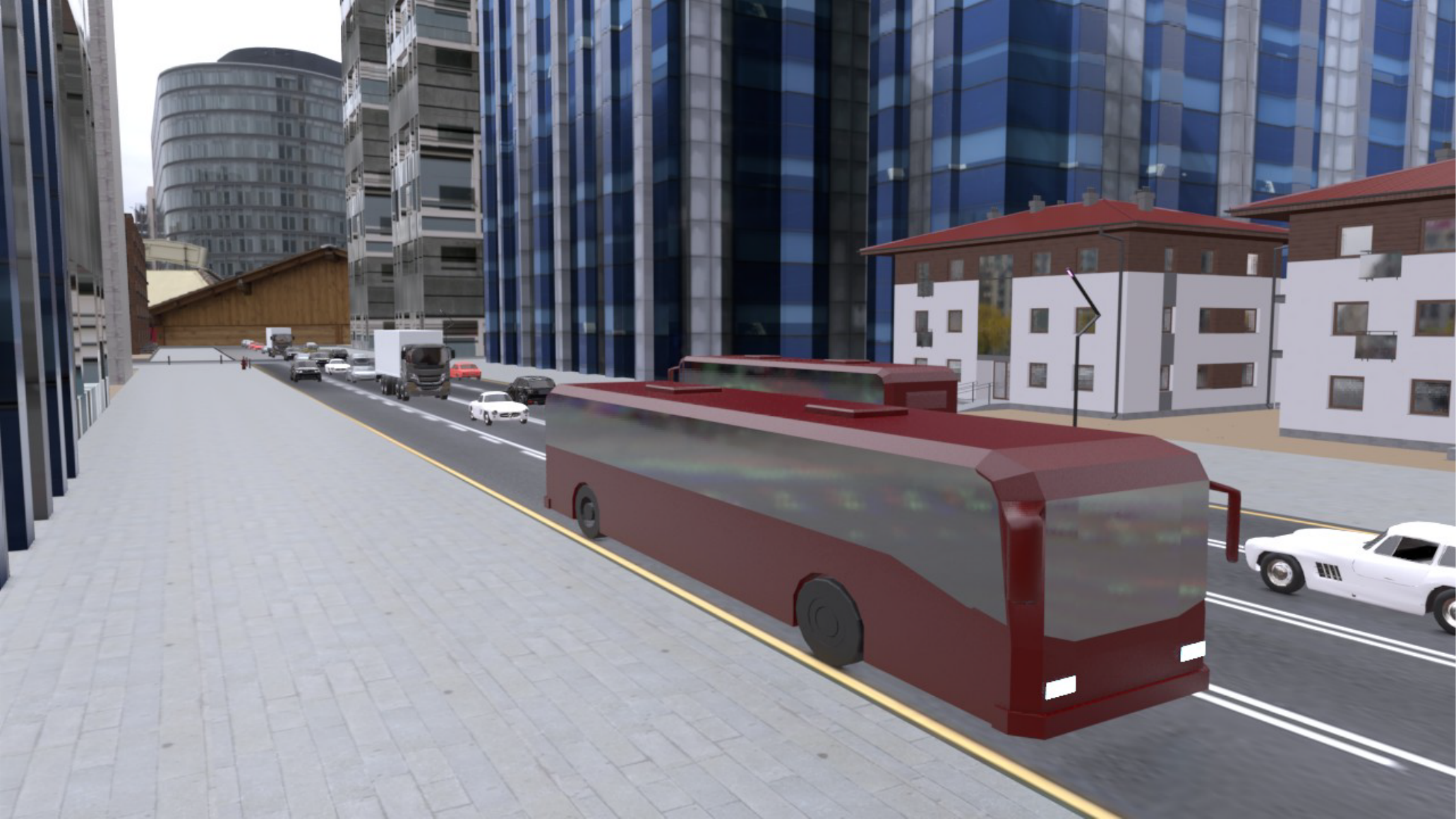}\label{}}
	\subfigure[Camera 6 of Basestation 2]{\includegraphics[width=0.48\textwidth]{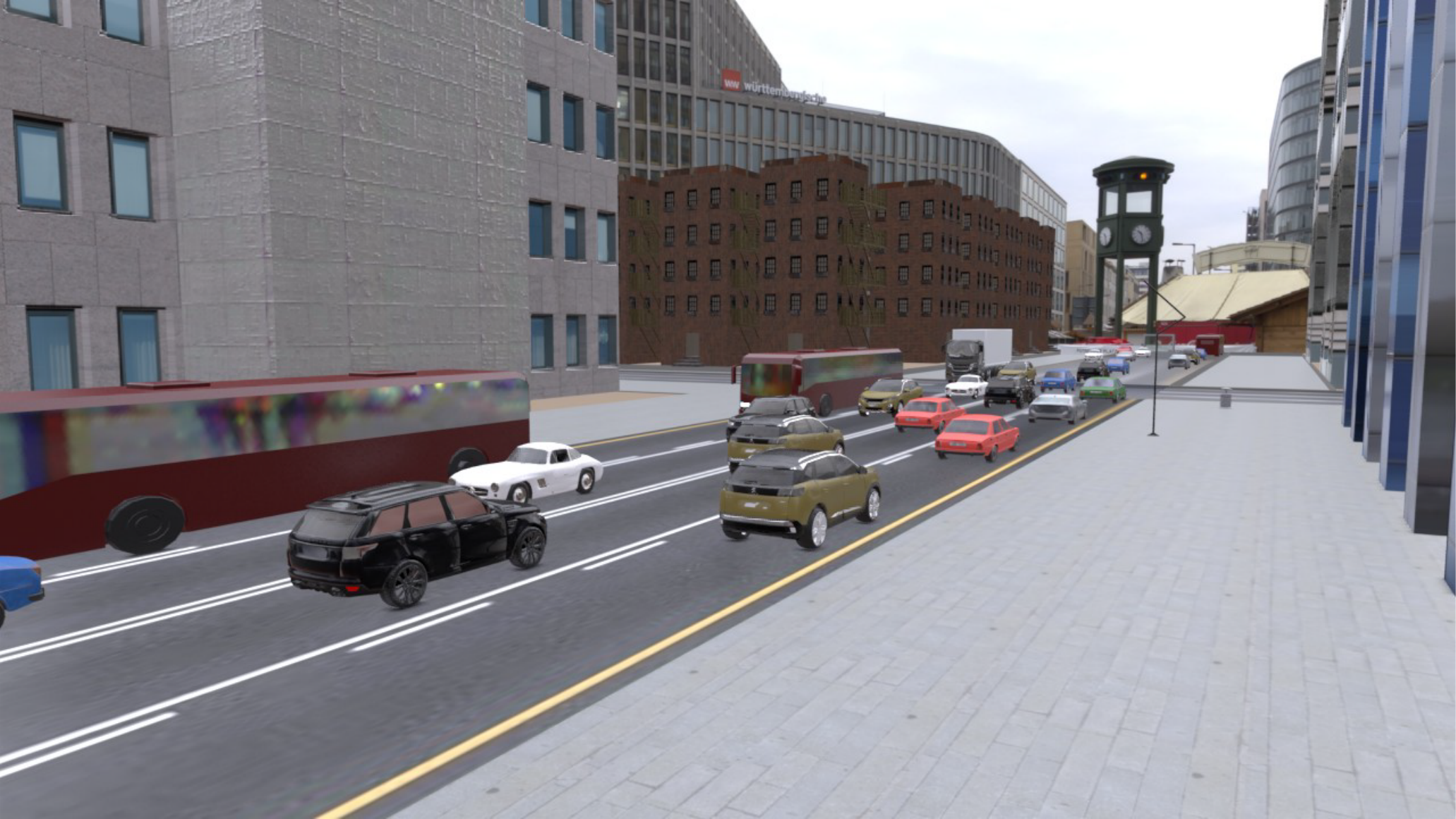} \label{}}
	\caption{This figure shows the far-peripheral view of the street captured from both the basestations. This figure presents the representative images captured by (a) Camera 3 of Basestation 1, and (b) Camera 6 of Basestation 2. The figure highlights the limited visibility issue encountered as the distance of the user from the basestation increases, making it even more challenging to predict the future blockage correctly.  }
	\label{ref:bs_images}
	\vspace{-4mm}
\end{figure}

\subsubsection{Camera-wise Analysis}
\begin{figure}[!t]
	\centering
	\includegraphics[width=0.6\linewidth]{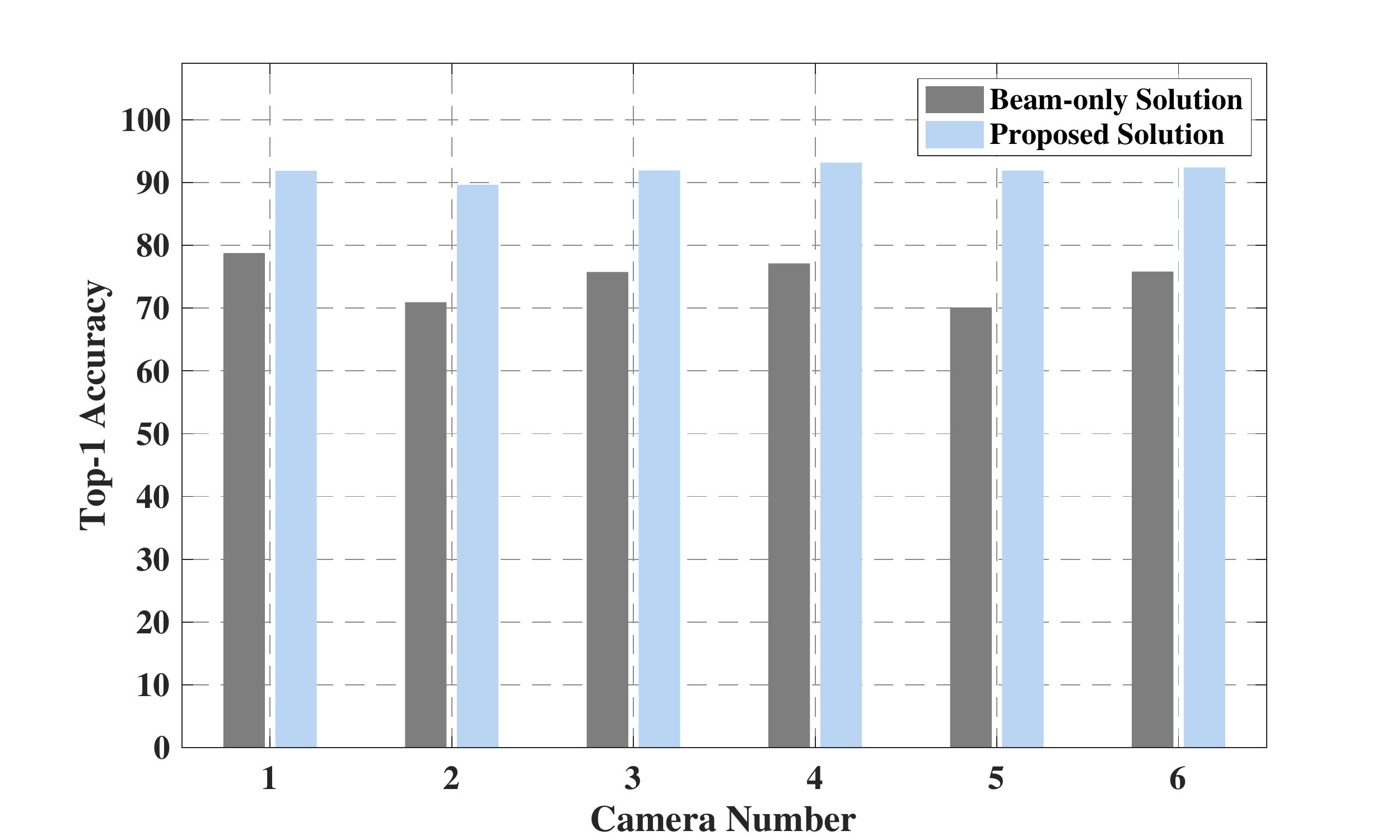}
	\caption{The figure presents the individual validation accuracy on the pivotal sequences for each camera for both the baseline (beam-only) solution and the proposed vision-aided approach. It is observed that the future blockage prediction performance of the proposed solution surpasses the prediction accuracy of the beam-only solution  for all the $6$ cameras, highlighting the efficacy of incorporating the visual component in the proposed solution. }
	\label{ref:bs_acc}
	\vspace{-4mm}
\end{figure}

There are a total of $6$ cameras capturing images in the dataset; each of them capturing different segments of the street. Cameras $1$, $3$ in basestation 1 and cameras $4$ and $6$ in basestation 2, capture the peripheral view of the street, see Fig.~\ref{ref:bs_images} for examples. In Fig.~\ref{ref:bs_acc}, we present the prediction accuracy per camera for the baseline and the proposed solutions. This accuracy is reported for the pivotal sequences only since they are the ones that present a clear challenge in a wireless network. The proposed deep architecture performs significantly better than the baseline approach for each camera. 
On average, the proposed solution exceeds the performance of the baseline solution by approximately $15\%$. 
However, an interesting thing to note here is the performance of the central cameras, i.e., camera $2$ and $5$. As compared to peripheral cameras, the prediction accuracy slightly degrades for the central cameras. The general expectation is that the solution will perform better for the central cameras as compared to the peripheral cameras. This is because vehicular motion in central cameras is parallel to the image plan making any displacement clearly visible. That dip in accuracy is due to the difference in data distribution between the cameras. Out of $\sim 4500$ samples in the validation set of cameras $2$ and $5$, approximately $1800$ samples are pivotal. Therefore, there is a slight data-imbalance for these two cameras, leading to a lower prediction accuracy as compared to the other cameras.

\subsubsection{Discussion on pivotal sequences}
In the previous two subsections, the performance of the proposed solution for future link blockage prediction was presented for all types of data samples. A common observation is that the proposed deep architecture achieves clear improvement over the baseline solution, especially for the pivotal samples; it achieves a blockage prediction accuracy of $90\%$ for \textbf{the pivotal sequences} in the validation set. Aside from the good performance of the proposed solution, we recognize that, in general, it fails in predicting blockages $10\%$ of the time, and, hence, we are seeking answers for why that is the case.
Unlike CNNs, it is difficult to study and visualize the reason behind the predictions of the recurrent component. 
Thus, any further improvement to the proposed solution warrants a detailed analysis of the failing cases. In particular, we analyze the architecture failures, and we present some probable causes of those failures. For this specific study, we only consider the pivotal sequences from the front cameras, i.e., camera $2$ and $5$, as they represent the main target of having a proactive blockage-prediction solution.

\begin{figure}[t]
	\centering
	\subfigure{	\centering \includegraphics[width=0.49\textwidth]{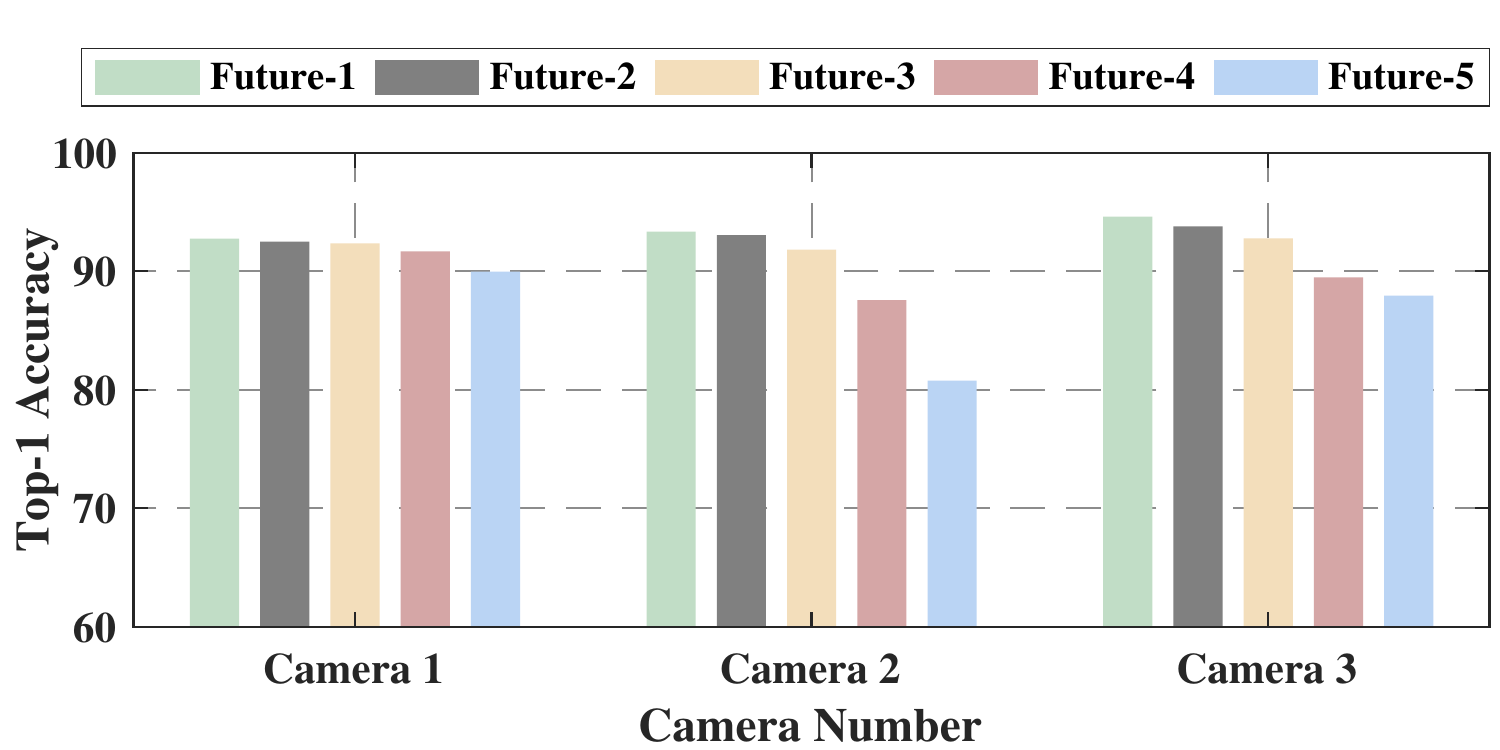}}
	\subfigure{\centering \includegraphics[width=0.49\textwidth]{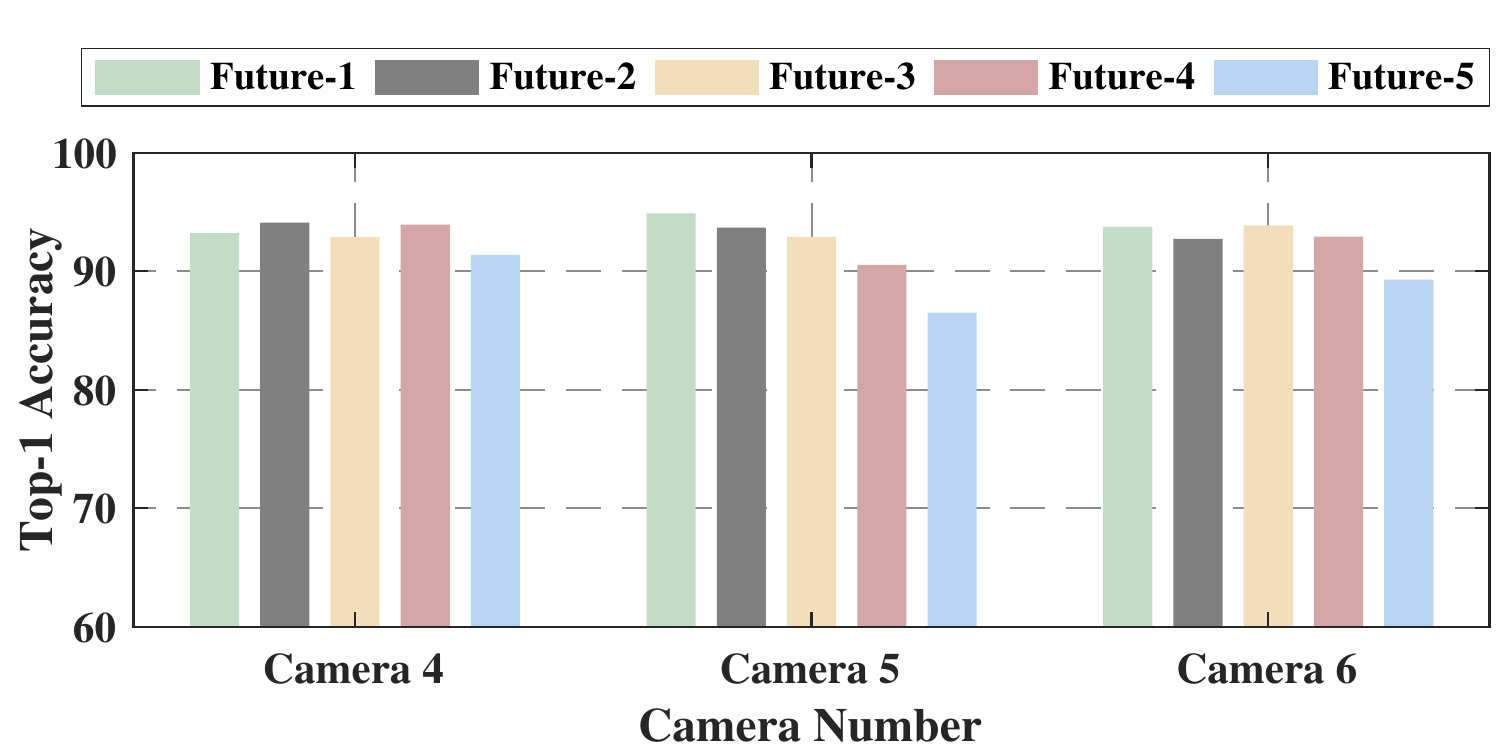}}
	\caption{The figure shows the impact of the future blockage instance on the proposed vision-aided model's prediction performance. The Top-1 accuracy versus the future blockage instance is presented for the cameras at the two basestations. It is observed that, generally, the further in future the blockage happens, the more difficult it is to predict the blockage.}
	\label{fig:acc_bar_plot}
	\vspace{-4mm}
\end{figure}

\begin{figure}[t]
	\centering
	\subfigure[Impact of user-blockage distance]{	\centering \includegraphics[width=1\textwidth]{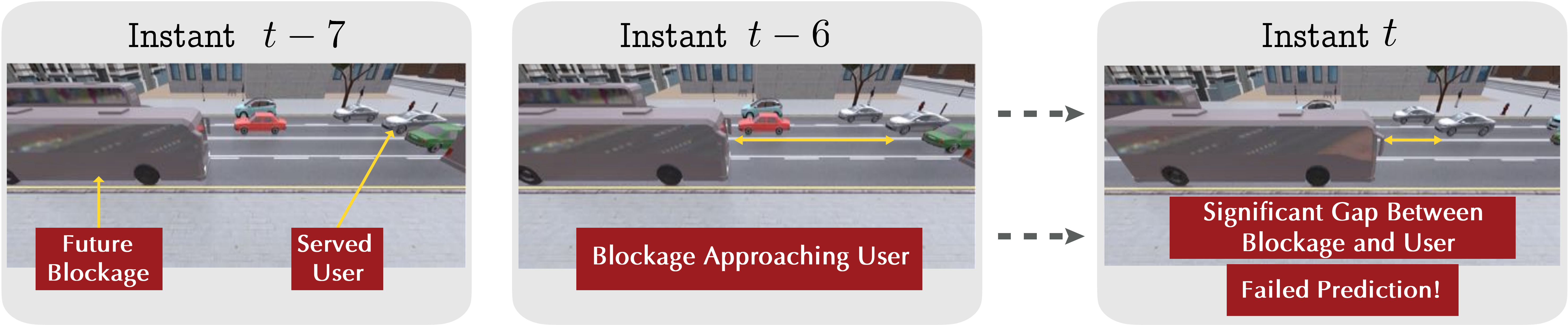}	\label{ref:failure_analysis_1}}
	\subfigure[Object detection failure]{   	\centering	\includegraphics[width=1\textwidth]{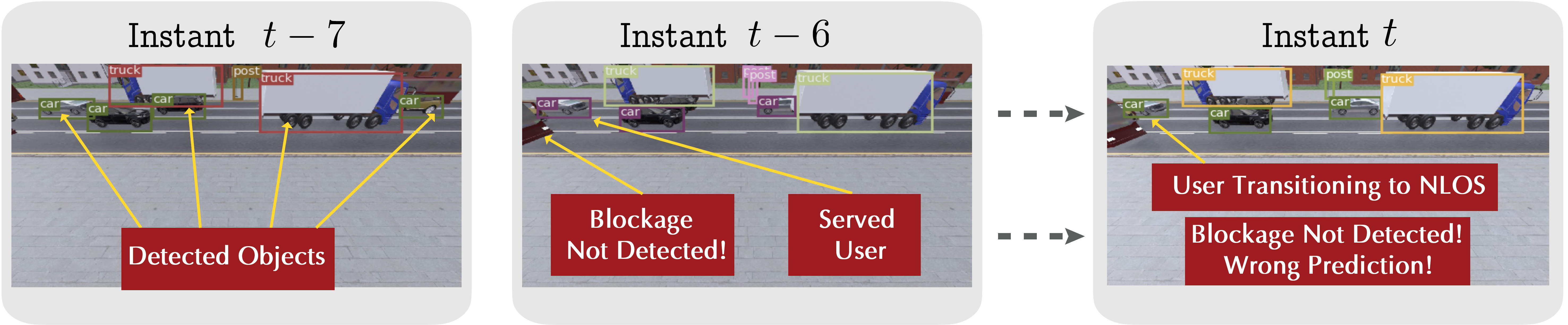}	\label{ref:failure_analysis_2}}
	\caption{This figure presents some illustration of the main reasons that may lead to failed predictions. As seen in figure (a) at time instant $t$, there is a significant gap between the blockage and the user, resulting in failed prediction. In figure (b), failure to detect the blockage bus resulted in failed future predictions. }
	\label{ref:failure_analysis}
	\vspace{-4mm}
\end{figure}

We identify two major reasons behind failed predictions, namely \textbf{instance at which blockage occurs} and \textbf{object detection failure}. As mentioned in Section~\ref{sec:blk_dataset}, we consider a sequence of $5$ future instances for generating the ground-truth link status of a user.
 We observe that the performance of the proposed solution varies with the instance at which the blockage occurs, e.g., blockage happens at first or fifth future instances. 
 In Fig.~\ref{fig:acc_bar_plot}, we present the top-1 validation accuracy versus the number of instances in the future window for all 3 cameras per basestation. The top-1 validation accuracy for the proposed solution increases as the blockage happens closer in time to the beginning of the future window. 
 In Fig.~\ref{ref:failure_analysis_1}, an example sequence is shown where the blockage happens at the $5$th future instance. It is observed that even at the $1$st future instance, there is a significant distance between the user and the blockage. Generally, the further we try to predict in the future, it becomes increasingly difficult for the deep neural network architecture to generalize. 

The second reason behind failed predictions could be traced back to the output of the object detector, i.e., the bounding-box values. If the object detector fails to detect a relevant object in any of the $8$ input sequences, it is very likely that the deep architecture is going to struggle. This is a consequence of the sequential nature of the proposed architecture.  In Fig.~\ref{ref:failure_analysis_2} an example of such case is highlighted. The object detector failed in detecting the incoming blockage, leading to a misprediction. This particular reason for failed predictions could be addressed through an end-to-end design, where blockage prediction is learned through training all components of a deep architecture together.

\subsection{Proactive Hand-off Prediction} 
As discussed in Sections \ref{sec:hand-off} and \ref{sec:hand-off_sol}, successful proactive hand-off can be viewed through the lens of blockage prediction. Two copies of the proposed architecture are deployed at basestation 1 and 2. The central unit receives the predictions for cameras 3 and 4, which have overlapping fields of view.
To evaluate the performance of proactive hand-off, we select a subset of sequences from the blockage prediction dataset. The subset is selected such that for each user, there are two conjugate sequences coming from basestations 1 and 2, i.e., the user gets blocked (NLOS) for one basestation and remains LOS for the other.
 Furthermore, the conjugate sequences in the subset are divided into two categories based on the serving basestation. The first category considers the case where a user is being served by basestation 1 and requires a hand-off to basestation 2 in the future, i.e., $z^{12}_u = 1$, while the second category is the opposite case, i.e., $z^{21}_u = 1$. 
The total number of conjugate sequences per category is $126$ for the first and $121$ for the second.

 \begin{table}[!t]
 	\caption{Proactive Hand-off}
 	\label{table}
 	\centering
 	\setlength{\tabcolsep}{5pt}
 	\renewcommand{\arraystretch}{1.4}
 	\begin{tabular}{|c|c|c|c|c|c|c|}
 		\hline
 		\multirow{3}{*}{\textbf{Model}} & \multicolumn{2}{c|}{\textbf{Hand-off Accuracy}}                                  & \multicolumn{4}{c|}{\textbf{Blockage Prediction Accuracy}}                                       \\ \cline{2-7} 
 		& \multirow{2}{*}{$n=1,\ n^{\prime} = 2$} &     \multirow{2}{*}{$n=2,\ n^{\prime} = 1$} &     \multicolumn{2}{c|}{Basestation 1} &     \multicolumn{2}{c|}{Basestation 2} \\ \cline{4-7}     
 		&                                      &                                          & \textbf{NLOS}        &     \textbf{LOS}       & \textbf{NLOS}        & \textbf{LOS}           \\ \hline
 		Beam-only                       & 73.41\%                                    & 72.67\%                                        & 77.25 \%               & 74.82 \%             & 78.81 \%                   & 76.24 \%             \\ \hline
 		Vision-aided Proposed Solution  & \textbf{86.50 \%}                       & \textbf{86.77 \%}                           & \textbf{92.86 \%}      & \textbf{90.90 \%}     & \textbf{91.73 \%}          & \textbf{90.47 \%}     \\ \hline
 	\end{tabular}
 	\label{ref:tab_handoff}
 	\vspace{-4mm}
 \end{table}
 
 In Table~\ref{ref:tab_handoff}, we present the performance of both the baseline and the proposed solutions for user hand-off. We present both the overall accuracy and the more fine-grained NLOS and LOS prediction accuracy for each scenario.
   The proposed solution achieves a hand-off prediction accuracy of $86.50 \%$ and $86.77 \%$, for category 1 and category 2, respectively, whereas the beam-only solution achieved $73.41 \%$ and $72.67 \%$ prediction accuracies. This is a reflection of the good proactive blockage prediction performance the deep architecture is capable of achieving. The last four columns of Table.\ref{ref:tab_handoff} confirms those results; they show the LOS and NLOS prediction accuracy per basestation. The proposed solution clearly achieves an approximately consistent improvement of $15\%$.

\section{Conclusion}
\label{sec:conc}
Enabling low latency and high reliability high-frequency wireless networks calls for the development of innovative solutions that overcome the challenges of LOS-link blockages. In this paper, we propose a bimodal machine learning solution capable of learning future link blockages in multi-user communication environments. It is based on deep neural networks, more specifically an object detector and a GRU network, and it relies on observing sequences of consecutive RGB frames and mmWave beams. The proposed solution is capable of proactively predicting blockages, and, hence, it enables mitigation measures to LOS blockages such as proactive user hand-off between basestations. This is demonstrated by developing and testing the proposed solution on a synthetic dataset of co-existing vision-wireless data, generated using the ViWi framework. The proposed solution achieves good blockage prediction performance, hitting an overall average test accuracy of approximately $88\%$, which goes up to $90\%$ when only pivotal sequences are considered. This accuracy reflects well on the user hand-off performance; a wireless network adopting the proposed solution at two different basestations is shown to approach a hand-off test accuracy of $87\%$. This does not only emphasize the importance of proactive blockage prediction for improving the reliability and latency performance of wireless networks but sheds some light on the role machine learning, and multi-modal data could play in shaping the future of those networks. 


\end{document}